\documentclass[a4paper]{mnras}

\usepackage{newtxtext,newtxmath}

\makeatletter

\newcommand{\Rmnum}[1]{\expandafter\@slowromancap\romannumeral #1@}
\makeatother

\usepackage{graphicx}	
\usepackage{amsmath}	
\usepackage{amssymb}	
\usepackage{subcaption}
\usepackage{color}

\def\apj{{\rm ApJ}}
\def\aj{{\rm AJ}}
\def\apjs{{\rm ApJS}}

\def\mnras{{\rm MNRAS}}
\def\cm{\, {\rm cm}}

\def\ergs{\, {\rm erg}\, {\rm s}^{-1}}
\def\angs{\, {\rm \AA }}
\def\asec{\, {\rm arcsec}}
\def\etal{{\rm et~al.\ }}
\def\mpc{\, {\rm Mpc}}
\def\kpc{\, {\rm kpc}}
\def\hmpc{h^{-1}{\rm Mpc}}

\def\msun{\, M_{\odot}}
\def\lya{Ly$\alpha$ }

\def\xiqe{${\xi}_{{\rm q} \alpha}$ }
\def\wqe{${w}_{{\rm q}\alpha}$ }
\def\xife{${\xi}_{{\rm f} \alpha}$ }
\def\xiqar{${\xi}_{{\rm q} \alpha}(r)$ }
\def\simlt{\lower.5ex\hbox{$\; \buildrel < \over \sim \;$}}
\def\simgt{\lower.5ex\hbox{$\; \buildrel > \over \sim \;$}}

\def\deltafibre{\delta_{\rm fibre}}

\def\ampqa{$b_{{\rm q}}b_{\alpha} f_{\beta} \langle \mu \rangle$}
\def\ampfa{$b_{{\rm f}}b_{\alpha} f_{\beta} \langle \mu \rangle$}

\newcommand{\df}{\delta_{\rm F}}

\title[Lyman-$\alpha$ intensity mapping]{
Intensity mapping with SDSS/BOSS Lyman-$\alpha$ emission, quasars
  and their
Lyman-$\alpha$ forest.
}

\author[R.A.C. Croft et al.]{\parbox{18cm}{
Rupert A.C. Croft$^{1}$\thanks{E-mail: rcroft@cmu.edu}
Jordi Miralda-Escud\'{e}$^{2,3}$,
Zheng Zheng$^{4}$,
Michael Blomqvist$^{5}$
and Matthew Pieri$^{5}$
}\vspace{0.3cm}\\
$^{1}$ McWilliams Center for Cosmology, Dept. of Physics, 
Carnegie   Mellon  University, Pittsburgh, PA 15213, USA\\
$^{2}$ Instituci\'{o} Catalana de Recerca i Estudis  Avan\c{c}ats, 
Barcelona, Catalonia\\
$^{3}$ Institut de Ci\`{e}ncies del Cosmos, Universitat de 
Barcelona/IEEC, Barcelona 08028, Catalonia\\
$^{4}$ Department of Physics and Astronomy, University of Utah, 115 S 1400 E, 
Salt Lake City, UT 84112, USA\\
$^{5}$A*MIDEX, Aix Marseille Universit\'{e}, CNRS,
 LAM (Laboratoire d'Astrophysique de Marseille) UMR 7326, Marseille, France
}

\begin{document}

\topmargin=-1.0cm

\maketitle


\begin{abstract}

We investigate the large-scale structure of \lya emission intensity in
the Universe at redshifts $z=2-3.5$ using cross-correlation
techniques. Our \lya emission samples are spectra of BOSS Luminous Red
Galaxies from Data Release 12 with the best fit model galaxies
subtracted. We cross-correlate the residual
flux in these spectra with BOSS quasars, and detect a positive signal
on scales $1\sim 15\hmpc$. We identify and remove a source of
contamination not previously accounted for, due to the effects of
quasar clustering on cross-fibre light.  Corrected, our quasar-\lya
emission cross-correlation is 50\% lower than that
seen by Croft et al.  for DR10, but still significant.  Because only
$\sim3\%$ of space is within $15 \hmpc$ of
a quasar, the result does not fully explore the global large-scale
structure of \lya emission. To do this, we cross-correlate with the
\lya\ forest. We find no signal in this case.  The 95\%
upper limit on the global \lya mean surface brightness from \lya\
emission-\lya forest cross correlation is $\langle \mu_{\alpha}
\rangle < 1.2 \times 10^{-22} \ergs\cm^{-2} \angs^{-1}\asec^{-2}$ This
null result rules out the scenario where the observed quasar-\lya\
emission cross-correlation is primarily due to the large scale
structure of star forming galaxies, Taken in combination, our results
suggest that \lya\ emitting galaxies contribute, 
but  quasars dominate within 
$15\hmpc$. A simple model for \lya emission from quasars based on
hydrodynamic simulations reproduces both the observed
forest-\lya emission and quasar-\lya emission signals.  The latter is
also consistent with extrapolation of observations of fluorescent
emission from smaller scales $r<1\hmpc$.

\end{abstract}


\section{Introduction}
\label{intro}

Intensity mapping (hereafter IM, Kovetz et al. 2016) 
refers to the use of one or more sharp spectral lines
to directly trace out the large-scale structure of the Universe from the line emission intensity in the three-dimensional space of angular coordinates and redshift, without
resolving discrete objects such as stars, galaxies or black holes. The
technique has been studied most prominently in the case of 21cm emission
from neutral hydrogen (e.g., Madau et al. 1997, Bandura et al. 2014 ), but theoretical predictions have been made
for various strong atomic and molecular lines, including hydrogen
\lya\ (e.g., Pullen et al. 2014), CO (e.g., Carilli, 2011) and 
CII (e.g., Pullen et al. 2018).
In the present paper
we investigate \lya\ intensity mapping, using spectra from the Sloan Digital Sky Survey (SDSS, Eisenstein et al. 2011) and 
theoretical predictions using cosmological hydrodynamic simulations.

Observational measurements of structure using line intensity are most easily
made using cross-correlation techniques, which avoids contamination by
interloper lines (Pullen et al. 2016). This can be done when some other tracer
of structure is available with a known redshift, such as galaxies or
quasars in a redshift survey. The first three dimensional IM result (
Chang et al. 2010)
 was obtained using 21cm data from the Green Bank telescope. A datacube of
angular position and wavelength (converted into redshift using the 21cm rest
wavelength) was cross-correlated with galaxy
angular positions and redshifts from the DEEP2 galaxy catalogue (Davis et al. 2001). The resulting
cross-correlation function of 21cm emission and galaxies was detected
on scales from 1 to $20 \hmpc$. A tentative measurement using CII line emission
has recently been made by Pullen et al. (2018), in cross-correlation with 
SDSS galaxies and quasars. The CO Power Spectrum Survey (Keating et al. 2015) has also made
a detection of fluctuations in CO emission, this time not using
cross-correlation techniques, but a $3 \sigma$ measurement of the 
bulk power spectrum. The final first detection we can mention 
at the present time is the quasar-\lya\ cross-correlation presented
by Croft et al (2016, hereafter C16) using both quasars and spectra from the SDSS tenth
data release (DR10). 
In that case, a \lya\ emission signal was seen within a separation of
$15 \hmpc$ from quasars, but with an amplitude that was larger than expected from
known \lya emitters. In this paper we revisit \lya\ IM using SDSS data, but 
this time we use DR12, which contains about 50\% more spectra, and more 
importantly we use a new tracer for measuring the cross-correlation, the \lya\  forest in absorption, alongside the quasars which were used before.

The hydrogen \lya\ line has a long history as a probe of star formation in the
high redshift Universe (Partridge \& Peebles, 1967). Its strength
 and position in the observed frame visible spectrum when produced
at the peak redshift for star formation has made it one of the prime
candidates for IM (Pullen et al. 2014 and references therein). Reprocessed radiation from quasars or
the UV background in general will also produce \lya\ emission (e.g., 
Gould and Weinberg, 1996), and these components have been considered
in IM predictions even at higher redshifts (Silva et al. 2013). Observations
of discrete \lya\ emitters from narrow band surveys and grism surveys, are now numerous
(e.g., Wold et al. 2017) and have been used to constrain the star formation rate observable
in \lya emission (e.g., Gronwall et al. 2007), and clustering of the star forming galaxies 
responsible (e.g., Guaita et al. 2010). More extended \lya emission has been seen as
low surface brightness halos around \lya emitting galaxies (Steidel \etal, 2011, Matsuda \etal, 2012, and
Momose \etal, 2014), and
the brightest extended objects have been seen proximate to quasars. These
bright \lya blobs (e.g., Cantalupo et al. 2014, Martin et al. 2014) have been studied using
Integral Field Unit (IFU) spectroscopy, revealing structure in
\lya emission on scales up to a few hundred kpc from the central quasar.

Although IFUs have been used to study \lya\ emission around quasars 
and galaxies (e.g., Gallego et al. 2018), the fields of view involved
are small, a few arcseconds across. For large scale structure studies, large
area spectral 
surveys are needed. There are several which are ongoing or upcoming,
including HETDEX (Blanc et al. 2011), PAU (Castander et al. 2012), J-PAS (Benitez et al. 2014) and SPHEREx (Dor\'{e} et al. 2014). Until the data is
available, one can use data taken for other purposes to do IM. In C16  
BOSS DR10 fibre spectra were used as \lya\ emission samples, and in the present
paper we will use BOSS DR12 (Alam et al. 2015). For cross-correlation one can use anything with
a known redshift in the desired range. As \lya\ is in the optical part of
the spectrum at redshifts $z>2$ the only large-area samples 
available for 
cross-correlation are quasars, and that is what was used in C16 and here
(in the present case, the BOSS DR12 quasar sample, P\^{a}ris et al., 2017). Quasars do allow us to use another tracer
of structure for cross-correlation, however, the \lya\ forest. This has
a low bias factor, leading to a lower signal in cross-correlation, but
it has the advantage that it traces the intergalactic medium over a whole line of sight in every spectrum, and including all the volume of the Universe instead of the small fraction
that is in the high density regions close to quasars.

As IM is carried out without identifying individual objects, contamination
from other sources, such as interlopers at different redshift, light
leakage, foreground or background emission is a difficult problem
which all IM experiments will have to deal with. With 21cm IM, the
Milky Way  acts as an extremely strong source of foreground emission which
is the limiting factor to current attempts to apply IM to the 
epoch of reionization.  Even with cross-correlation techniques
it is easy for  contamination to affect a potential signal, 
as was found in C16 (see appendices in that paper), where the cross-talk effect among
spectra from fibres that are placed nearby in the BOSS camera implies that
the \lya emission light from quasars can pollute spectra in adjacent fibres
used for \lya detection, introducing artificial cross-correlations. These
fibre pairs placed close to each other in the BOSS CCD camera were excluded from the
cross-correlation, but as we shall see an effect due to quasar clustering that was
neglected turns out to be important. In general, the subtleties of 
light contamination remain to be explored fully for precise measurements.
One of the purposes of the present study is to reveal how contamination
can enter and could be mitigated by properly designed dedicated IM  experiments. 

After dealing with contamination, C16 found the suprising result
that there appears to be a high \lya\ surface brightness around 
quasars at redshifts $z=2-3.5$.
It was deemed most likely that this was due to star forming galaxies,
although this necessitated that most \lya\ photons emitted by such galaxies
are visible, but have not been detected by other means (perhaps due
to an extremely low surface brightness). The other possibility was that the
energy emanating from QSOs was the source of the \lya\ emission instead. We will aim
in this paper to decide between these two models, using a new dataset
which includes the \lya\ forest.

Whether quasars or galaxies are the source of the quasar-\lya\ emission
cross-correlation signal seen in C16, it is not clear whether a physical
model can be constructed in the context of the large-scale distribution
of matter and quasars predicted by the Cold Dark Matter model which
is consistent with observations. We will set up some toy models
based on cosmological hydrodynamic simulations to address this. There
are many sophisticated models of \lya\ emission  and large-scale
structure which use \lya\ radiative transfer (e.g., Smith \etal 2018, Kakiichi \& Djikstra 2017, 
Zheng et al. 2010, Kollmeier \etal 2010), but we will not be modeling the physical processes 
associated with \lya\ emission, scattering and absorption. Instead we
will "paint" a \lya\ emission field onto the large-scale structure in 
the simulations and see whether it yields a quasar-\lya\ emission
cross-correlation and a \lya\ emission- \lya forest cross-correlation
that are consistent with the measurements within the observational errors.

Our plan for the paper is as follows. In Section 2 we introduce the observational
data we will be using, SDSS LRG spectra, quasars, and \lya\ forest spectra. We compute the quasar-\lya emission cross-correlation in Section 3 including
comparison to the linear CDM model. We also measure the projected cross-correlation
function and compare to an extrapolation of existing data on small scales. In 
Section 4 we focus on the \lya forest-\lya emission cross-correlation, attempting a
measurement from SDSS data and using it to constraining the global \lya surface
brightness. In Section 5 we describe some toy models for the \lya emission intensity
based on cosmological hydrodynamic simulations and then compare them to our 
observational results. In Section 6 we summarize our findings and discuss their
implications. In an Appendix, we show how we minimize light contamination in the 
measurements and test our methods with mock observations. 

\section{Observational Data}

This study makes use of data from the SDSS BOSS survey Data Release 12
(DR12), including quasar
position and redshift data, and galaxy and quasar spectra.
 The SDSS camera and telescope
are  described in Gunn et al. (1998) and Gunn et al. (2006), respectively.
Full information on the SDSS/BOSS spectrographs can be found in
Smee et al. (2013). The wavelength coverage of the spectrograph is from
$\lambda=$3560 \AA\ to 10400 \AA\, the resolving power is $R\sim1400$
for the range $\lambda= 3800$ \AA$-$ 4900 \AA, and is kept above
$R=1000$ for the remainder of the wavelength range. The
fibres have a diameter of 120 $\mu$m, corresponding to $2$ arcsec in angle.
We restrict the redshift range of data we use in our analysis to
$2.0 < z < 3.5$, due to the spectrograph cutoff at low redshift and
the limited number of observed quasars at high redshift.

\subsection{Quasar catalogue}

\label{qcat}

For cross-correlation, we use quasars from the DR12 catalogue
(P\^{a}ris
et al. 2017), which contains 297,301 spectroscopically confirmed quasars.
The spectroscopic target selection
procedure is that of Ross et al. (2012), which
combines several algorithms to identify candidates, individually 
described in Richards et al. (2009); Kirkpatrick
et al. (2011); Y\'{e}che et al. (2010); Bovy et al. (2011).
The quasar redshifts have been obtained using the Principal Component
Analysis method described in Paris et al.
(2012), and Paris et al. (2017).
We apply a quasar redshift cut, selecting only objects within the range
$z=2.0-3.5$. This reduces the number of quasars we use to 218726, with a 
mean redshift of $z=2.55$.

\subsection{Quasar \lya\ forest spectra}

We also use quasar \lya\ forest spectra in our cross-correlation
studies with \lya\ emission. The quasar spectra in our sample are
selected from the DR12 quasar catalogue. In order
to avoid very short spectrum lengths,  we use a slightly tighter cut
of $z=2.05-3.5$ than for the quasars in Section \ref{qcat}. 
 The BAL quasars are discarded, as well as 3188 (very noisy) spectra which are flagged and removed by requiring a median S/N $>0$, normalisation factor $>0$, a minimum  of 75 spectral pixels in the spectrum and a successful continuum fit (see below).
These steps lead to a set of 161213 quasars in the forest
sample. 

We use the public DR13 pipeline reductions (Albareti et
al., 2017) of the spectra, selecting the \lya\ forest
pixels which are in the restframe wavelength range of 1040-1200 \AA\ .
We apply a minimum observed wavelength range of 3600 \AA\, and mask
pixels affected by strong skylines based on the DR12 sky mask. We also mask
DLAs and correct for the DLA wings, following the method described in Lee et al. (2013), using the catalogue of Noterdaeme et al. (2018) \footnote{\tt http://www2.iap.fr/users/noterdae/DLA/DLA.html}, 
(the method is described in Noterdaeme et al., 2009, 2012)
to identify them. We fit the continuum to each spectrum
 using "method 1" of Busca et al. (2013) and (also known as "C1" by
Delubac et al., 2015). To speed up the analysis we combine three adjacent 
spectral pixels into analysis pixels. The total number of these 
\lya\ forest pixels used in our cross-correlation analyses is
22.92  million, and their mean redshift is $z=2.41$.

\subsection{Galaxy spectra}
 The 1570095 galaxy spectra in our sample are of targeted LRGs which are within
redshifts $z\sim0.15$ and $z\sim 0.7$ (mean redshift $z=0.48$). In our study, as in C16, these galaxy spectra are used to measure \lya\ emission, after a model spectrum of the targeted galaxy is subtracted from the observed spectrum.
The redshift range of the LRGs is not
important for our purpose. For each spectrum, we make use only of the pixels
for which the \lya emission line is within the redshift range
specified above ( $2.0 < z < 3.5$). In observed
 wavelength this is from 3647 \AA\ to 5470 \AA. 

The  BOSS LRG program (Dawson et al. 2013)  targeted two galaxy
 samples,  CMASS
("constant mass") and LOWZ ("low-redshift"). The
LOWZ galaxy sample is composed of massive red galaxies
and spans $0.15 \simlt z  \simlt 0.4.$ The CMASS
 sample spans $0.4 \simlt z \simlt 0.7$. 

The faintest galaxies are at $r = 19.5$ for LOWZ and $i = 19.9$ for
CMASS. Both samples are color-selected
to provide near-uniform sampling over the combined volume.
 Colors and magnitudes for galaxy selection
are corrected for Galactic extinction using Schlegel et al. (1998)
dust maps. We do not differentiate between the samples
in our analysis.   

The spectroscopic measurement pipeline for BOSS is described in detail in 
Bolton \etal (2012). The data products that are used in the 
present analysis are:
(a) wavelength-calibrated, sky-subtracted, flux-calibrated, and
co-added object spectra, which have been rebinned onto a uniform baseline
of $\Delta \log_{10} \lambda = 10^{-4}$ (about 69 km s$^{-1}$ pixel$^{-1}$),
(b) mask vectors for each spectrum, and
(c) statistical error-estimate vectors for each spectrum (expressed
as inverse variance) incorporating contributions
from photon noise, CCD read noise, and sky-subtraction
error.

\subsection{Data preparation}

Our \lya\ emission samples are LRG spectra with the galaxy spectrum subtracted.
The galaxy spectrum in each case is the best fit model
provided by the pipeline.
This template model spectrum (see Bolton \etal 2012 for details)
is computed using least-squares minimization comparison
of each galaxy spectrum to a full range of 
galaxy templates. Smooth terms absorb Galactic extinction, intrinsic
extinction, and residual spectrophotometric calibration errors
(typically at the $10\%$ level) that are not fully spanned by
the template basis sets.
These basis sets are derived from restframe
principal-component analyses (PCA) of training samples
of galaxies.

After this subtraction, we compute the mean residual surface
brightness per unit observed wavelength from all the spectra. We subtract this from each
spectrum, as we are only interesting in fluctuations in the \lya\ emission
surface brightness. This procedure is done as in C16, where it is describe in more detail. We also reject any pixels which fall within the mask
for sky lines, as we have done with the \lya\ forest spectra.

\section{Quasar-Lyman-alpha emission cross-correlation}
\label{secxiqe}

Before computing the quasar-\lya emission cross-correlation, we first
split the sample of galaxy and quasar spectra into 160 subsamples of
approximately equal sky area based on contiguous groupings of plates.
We then convert the galaxy spectrum pixels and the quasar angular positions
and redshifts into comoving Cartesian coordinates using a flat cosmological
model with matter density $\Omega_{\rm m}=0.315$, consistent with the
Planck, Ade (2014) results (cosmological constant
density $\Omega_{\rm \Lambda}=0.685$). This fiducial model is used throughout
the paper.

We compute the  quasar-\lya emission surface brightness
cross-correlation, \xiqar , using a sum over all
quasar-galaxy spectrum pixel pairs separated by $r$
within a certain bin:
\begin{equation}                                                               
  \xi_{{\rm q} \alpha}(r) = \frac{1}{\sum_{i=1}^{N(r)}w_{ri}}                  
  \sum^{N(r)}_{i=1} w_{ri}\, \Delta_{\mu,ri},       
                  \label{xieq}                                        
\end{equation}
where $N(r)$ is the number of pixels in the
bin centered on quasar-pixel distance $r$,
and $\Delta_{\mu,ri}=\mu_{ri}-\langle \mu(z) \rangle$
is the residual surface brightness
in the spectrum at pixel $i$ for the bin $r$.
Note here that we have a different list of pixels labeled as $i$ for
each bin in the separation $r$ between a pixel and a quasar, which has
\lya surface brightness
$\mu_{ri}$. The residual flux at each pixel is obtained by
subtracting the mean at each redshift, $ \langle \mu(z) \rangle$.
We weight each pixel  by $w_{ri} = 1/\sigma^{2}_{ri}$, where
$\sigma^2_{ri}$ is the pipeline estimate of the inverse variance of the
flux at each pixel. We first present our results as a function of
only the modulus of the quasar-pixel separation $r$ in comoving $\hmpc$,
 in 20 bins logarithmically spaced between $r=0.5 \hmpc$ and $r=150 \hmpc$.

In C16 it was shown that stray light from the quasars themselves 
could contaminate the spectra of galaxies which are nearby on the 
spectrograph CCD. This
occurs because the
light from the various fibres is dispersed onto a single CCD, so that
extraction of each spectrum along one dimension (Bolton \etal 2012)
may include light from
adjacent fibres on bright sources. Tests were carried out in C16
showing the presence of this
stray light from
quasars in galaxy spectra when the quasar and galaxy spectra were four
fibres apart or fewer, in the list of fibres as they are ordered in
the CCD. A constraint was applied so that galaxy and quasar spectra should
be at least six fibres apart when computing the flux cross-correlation of
equation \ref{xieq}. We have applied this restriction in this paper, both
for the quasar-\lya\ emission cross-correlation and for 
the \lya\ forest-emission
cross-correlation (presented later in Section \ref{secxife}).

 In this paper, we have also 
carried out a further examination of systematic errors, and have found 
that this cross-fibre light can affect the results
even when excluding close fibres from the cross-correlation. This is due
to quasar clustering. Even when a specific quasar-galaxy pair is
not used in the cross-correlation, light from other quasars spatially 
clustered with the first can contaminate nearby galaxy fibres. 
In Appendix A, we have tested three different methods for eliminating this
contamination, which give statistically consistent results. Doing this
involves removing all pixels which might be problematic from
the dataset entirely (rather than just exclusing certain pairs from
the cross-correlation). This necessarily reduces the statistical
power of the dataset. For example, removing all galaxy spectra
which have a $z=2.0-3.5$ quasar within $\pm 5$ fibres reduces the number of
galaxy spectra used from 1570095 to 37\% of that number (574603).
 After accounting for the additional
source of systematic error, we find a lower quasar-\lya\ cross-correlation
function amplitude (by a factor of approximately 2) compared to C16 (see 
Section \ref{modelfit} below for model fitting amplitude). 
For details of the cross-fibre
light contamination, the reader is referred to Appendix A.

\begin{figure}
\centerline{
\includegraphics[width=0.45\textwidth]{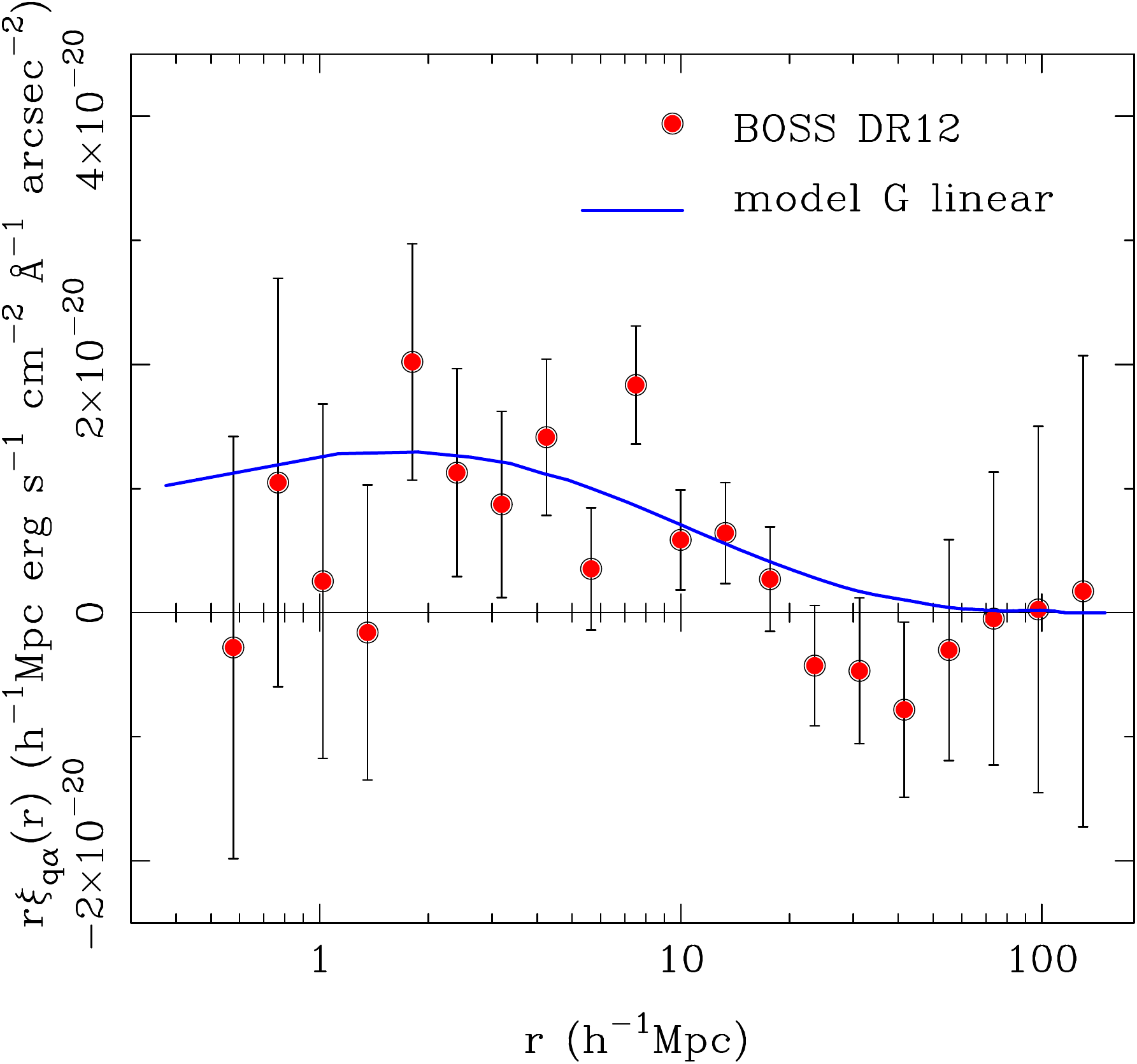}
}
\caption{ The quasar-\lya emission cross-correlation function,                 \xiqar\ (see Equation \ref{xieq}). The points represent                        results for the fiducial sample that covers redshift range  $2.0 < z< 3.5$.    The error bars have been calculated using a jackknife estimator and            160 subsamples of the data. The smooth curve is a best fit linear CDM          correlation function (see Section \ref{modelfit}).
  \label{cdmfid}}
\end{figure}

\begin{figure}
\centerline{
\includegraphics[width=0.45\textwidth]{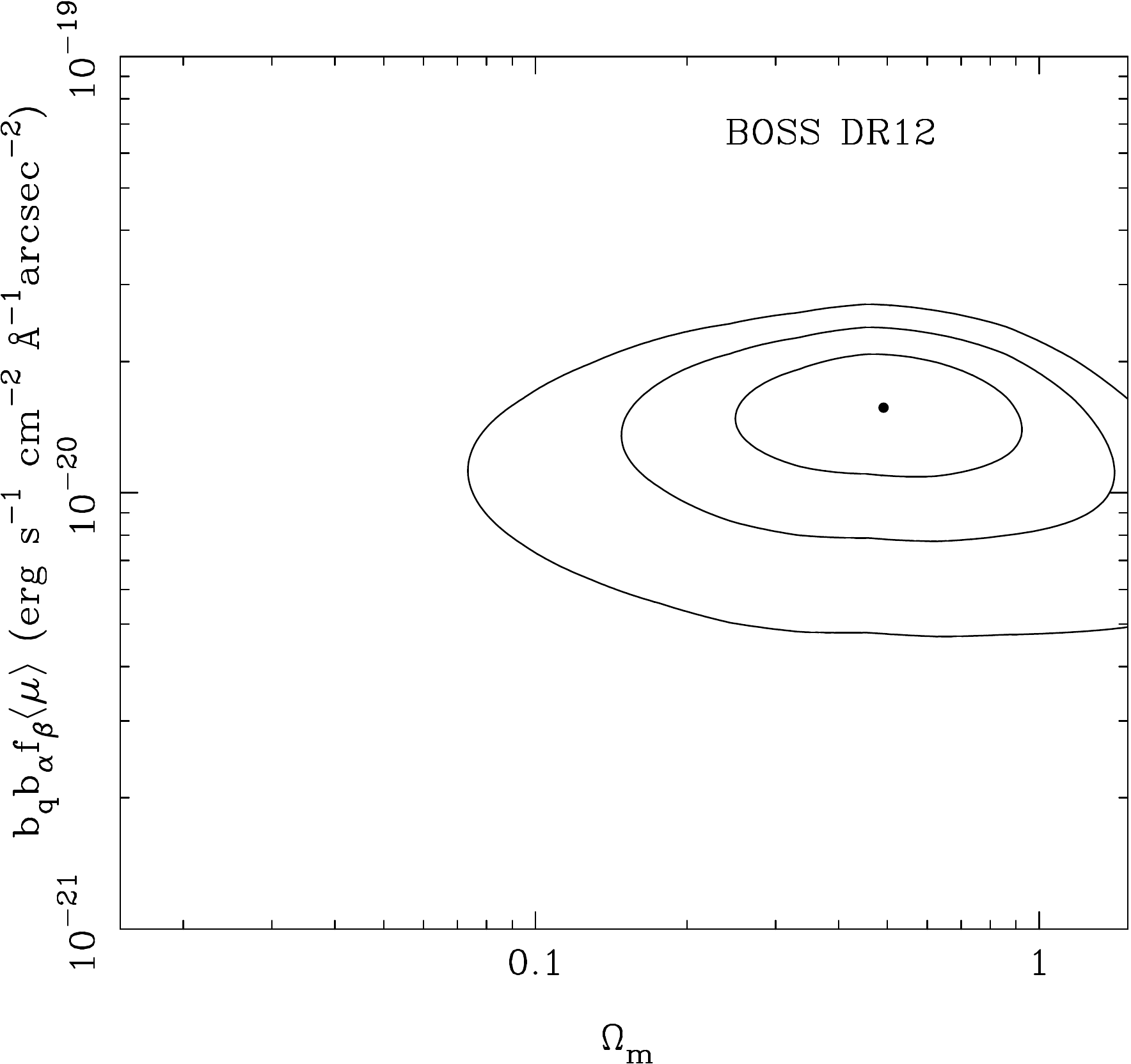}
}
\caption{ 
Fit parameters for the amplitude   \ampqa\  and shape $\Omega_{\rm m}$ (for fixed $h$ and other parameters)        
of a linearly biased CDM model fit to the \lya cross-correlation                
function plotted in Figure \ref{cdmfid}. The dot indicates the best fit         
parameters and the contours show the 1, 2 and 3 $\sigma$ confidence contours.
  \label{fitfid}}
\end{figure}

We  use the 160 Jackknife samples to compute the mean 
 of \xiqar\ and the error on the mean, as well as the covariance
matrix of \xiqar\ (as in C16, Equation 7). 
In Figure \ref{cdmfid} we show  \xiqar\ for our 
fiducial sample (which is the entire
dataset over the redshift range $2.0 < z< 3.5$. As in C16, the mean redshift
of  the sample is $z=2.55$. We have removed the cross-fibre light 
contamination from the results using the pixel exclusion method from Appendix 
A.  We can see from Figure \ref{cdmfid}
that there is evidence for an excess \lya\ surface brightness within
$r\simeq 15 \hmpc$ of quasars, but not beyond this.

\begin{figure}
\centerline{
\includegraphics[width=0.45\textwidth]{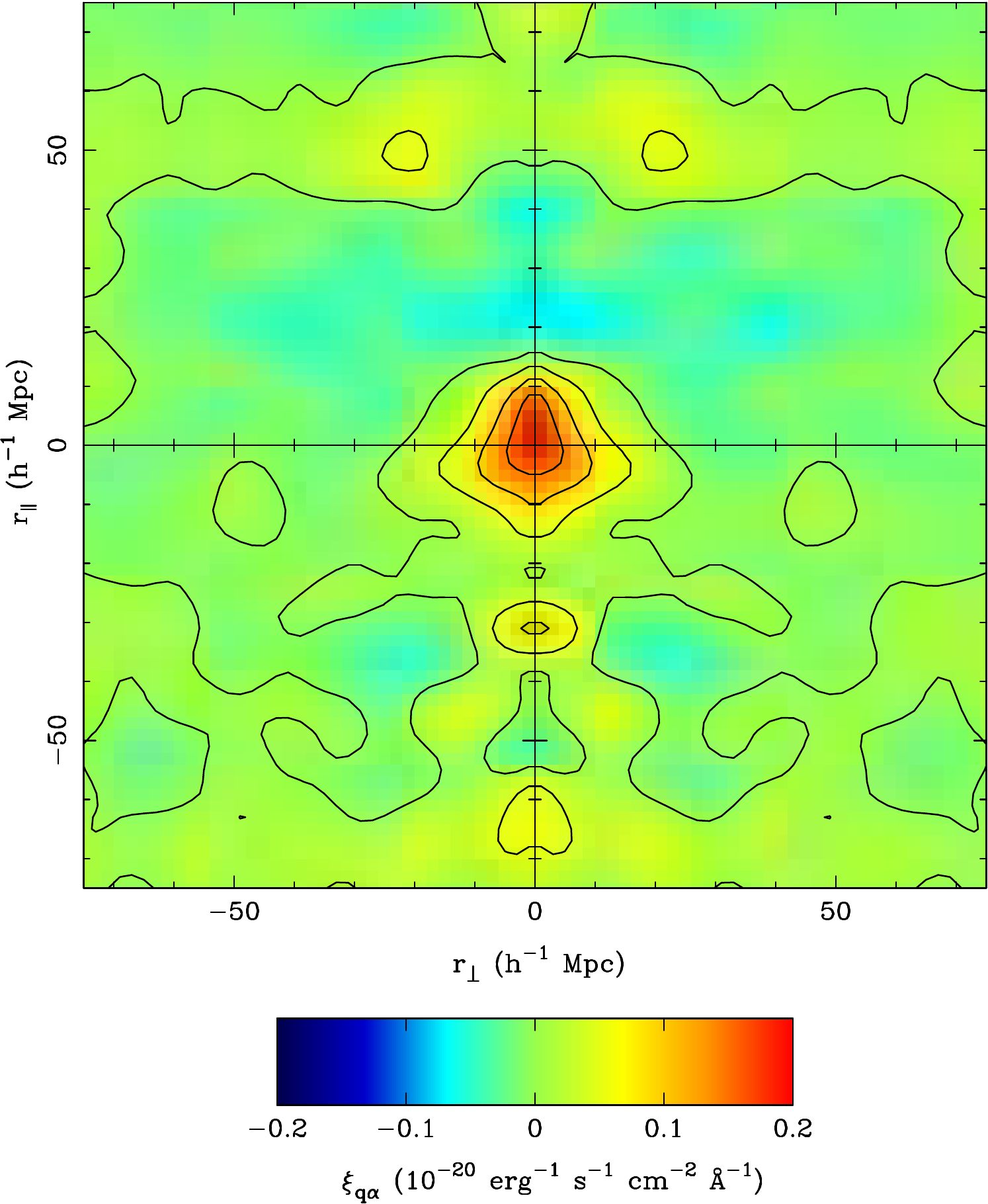}
}
\caption{ 
The quasar-\lya cross-correlation \xiqe\ as a function             
of $r_{\parallel}$ and $r_{\perp}$. The units (of \lya                         
surface brightness) are the same as in Figure                                  
\ref{cdmfid}. The contours are spaced at values of $10^{-21} \ergs             
\cm^{-2} \angs^{-1} \asec^{-2}$. To reduce noise in                            
the image, the dataset was smoothed with a Gaussian filter with                
$\sigma=4 \hmpc$ before plotting. Light contamination was excluded by
using pixel veto (see Section \ref{removecontam} of Appendix A). 
\label{sigpi}}
\end{figure}

We also visually examine redshift space
anisotropies in the
correlation function \xiqe\ by considering bins in the parallel and
perpendicular components of $r$, $r_{\parallel}$ and $r_{\perp}$. This is
plotted in Figure \ref{sigpi}. We can see that there appears to 
be some sign of
 visual elongation of the inner contours in the line of sight direction. 
In C16, the apparent signal to noise of the detection was large enough
that we were able to make a measurement of this elongation, noting 
that it was consistent with that expected due to radiative transfer effects 
(Zheng et al., 2011). In the present work, due to the additional data cuts 
necessary to avoid all light contamination, the significance of the 
signal is lower and we do not attempt to quantify any anisotropy, leaving
this for future work with larger datasets.

\subsection{Linear CDM fit: model G}

\label{modelfit}

In Figure \ref{cdmfid}, a solid curve is plotted. This is a Cold
Dark Matter model
fit. 
If the \lya emission clustering is due to star forming galaxies tracing
 a linearly biased
version of the density field, then a model for the isotropically
averaged quasar-\lya cross-correlation \xiqar is (see C16 for more details):
\begin{equation}                                                                
\xi_{{\rm q}\alpha}(r)= b_{q}b_{\alpha} f_{\beta} \langle \mu_{\alpha} \rangle \
\xi(r)                                                                          
\label{model}                                                                   
\end{equation}
where $\langle \mu_{\alpha} \rangle$ is the mean surface
brightness of \lya emission, $b_{\rm q}$ and $b_{\alpha}$ are the
quasar and \lya emission linear bias factors,
$\xi(r)$ is the linear $\Lambda$CDM mass correlation function,
and $f_{\beta}$ is a constant enhancement to the correlation function on
linear scales that
is caused by peculiar velocity redshift-space distortions (Kaiser 1987).
We refer to this model where galaxies are the source of \lya emission
as model G. Equation \ref{model} represents the linear theory
version of model G. In Section \ref{secsim} we examine a version of model
G based on simulations, as well as model Q, where quasars are responsible
for the \lya emission seen in the \xiqe signal.

As in C16 we carry out a $\chi^{2}$ fit for two free parameters in 
model G, the 
amplitude $\text{\ampqa}$
and the shape, parametrised in the CDM power spectrum by $\Omega_{\rm m}$.
In Figure \ref{fitfid} we show the best fit values of these parameters
and their confidence contours. The best
fit values and their one dimensional marginalized $1\sigma$ error
intervals are
\begin{equation}                                                               
\text{\ampqa}  = 1.60^{+0.32}_{-0.30} \times 10^{-20} \ergs \cm^{-2} \angs^{-1}
\asec^{-2},                       
                          \end{equation}
and $\Omega_{m}=0.491^{+0.235}_{-0.174}$.
The shape,
 $\Omega$ is 1 $\sigma$ higher from the
value obtained by C16, but the amplitude is lower by a factor of 2. These
differences result from the removal of a newly discovered source of 
light contamination in the present work, due to  (see Appendix A). Qualitively, however,
the conclusions drawn by C16 about \xiqe remain applicable here: first
that \xiqe is
consistent with the linear version of model G, but there is
only clustering detected on scales smaller than $15 \hmpc$, and second
that the amplitude in this model is much
higher than would be expected if the \lya emission detected were due
to known \lya emitters (see C16, Section 5 for a detailed treatment).
We briefly recap here how these conclusions are reached.

In the context of this model (G), we take the amplitude \ampqa and use the
published quasar bias value $b_{q}=3.64^{+0.13}_{-0.15}$
(Font-Ribera \etal, 2013), and the $f_{\beta}=0.8\pm0.15$ value from
C16, 
to compute the mean \lya surface brightness
at $z=2.55$, finding
\begin{equation}                                                                
\langle \mu_{\alpha} \rangle=(1.9 \pm 0.5) \times10^{-21}                       
(3/b_{\alpha}) \ergs\cm^{-2} \angs^{-1}\asec^{-2} ~.                            
\label{eq:backi}                                                                
\end{equation}
here $b_{\alpha}$ is the luminosity weighted
 bias factor of \lya emission from star forming
galaxies at $z=2.55$, which C16 estimate should be $b_{\alpha} \sim 3$.

We convert this into a comoving
\lya  luminosity density $\epsilon_{\alpha}$ using
\begin{equation}                                                                
\epsilon_{\alpha}=4\pi \langle\mu_\alpha \rangle                                
                  \frac{H(z)}{c}\lambda_{\alpha}(1+z)^{2},                      
\label{eq:backe}                                                                
\end{equation}
where $c$ is the speed of light and $\lambda_{\alpha}=1216 \angs $.
We find the value $\epsilon_{\alpha}=1.5\times10^{41}  (3/b_{\alpha})           
\ergs \mpc^{-3}$.
We then use the 
relationship ${\rm SFR}/(\msun\,  {\rm yr}^{-1})=L_{\alpha}/(1.1\times10^{42} \ergs)$ (Cassata et al. 2011)
 to convert this into a measurement
of the star formation rate density:
\begin{equation}                                                                
\label{sfrfid}                                                                  
{\rho}_{\rm SFR}(z=2.55) = (0.14\pm0.04) {3\over b_{\alpha} }                   
 \msun\, {\rm yr}^{-1} \mpc^{-3} ~.                                             
\end{equation}

This value is  15 times  higher (for $b_{\alpha}=3$) than the 
\lya emitter based measurements of Gronwall \etal (2007), Ouchi \etal (2008)
or Cassata \etal (2011). As discussed in C16, this seems unlikely, but 
in absence of other constraints would be possible, as the extinction-corrected
star formation rate density is similar to this value (Bouwens \etal, 2010). 
In the present paper, however we shall 
show that there are two pieces of evidence
that model G with this high value of \lya\ luminosity density does
not apply to our Universe, but instead that quasars themselves are likely
to be responsible for the \xiqe signal. The first piece of evidence 
involves   comparison with the \lya emission seen around quasars on
smaller scales, in the next Section.

\subsection{Projected quasar-Lyman-alpha emission}

In recent years, there have been several successful searches for \lya emitting
nebulae in close proximity to quasars. Cantalupo et al. (2014) and
Hennawi et al. (2015) used custom narrow band filters to find
two bright nebulae with diameters
460 proper $\kpc$ and 350 $\kpc$ respectively 
around quasars at redshifts $z\sim2$. Martin \etal (2014) used the Cosmic
Web Imager, an integral field spectrograph to detect extended
emission around a large \lya blob centered on a quasar. 
In these three cases the emission was consistent
with quasar induced fluorescent emission (Hogan \& Weymann 1987, 
Gould \& Weinberg 1996, Cantalupo et al. 2005, Kollmeier et al. 2010). Although the detection
rate with narrow band imaging was low (ten percent), a large IFU survey
with the MUSE Spectrograph by Borisova \etal (2016) found that large
\lya nebulae appear to be ubiquitous around bright radio-quiet quasars,
with 17 examples found. These observations were stacked by Borisova \etal,
yielding the circularly averaged surface brightness  profile around quasars.

\begin{figure}
\centerline{
\includegraphics[width=0.45\textwidth]{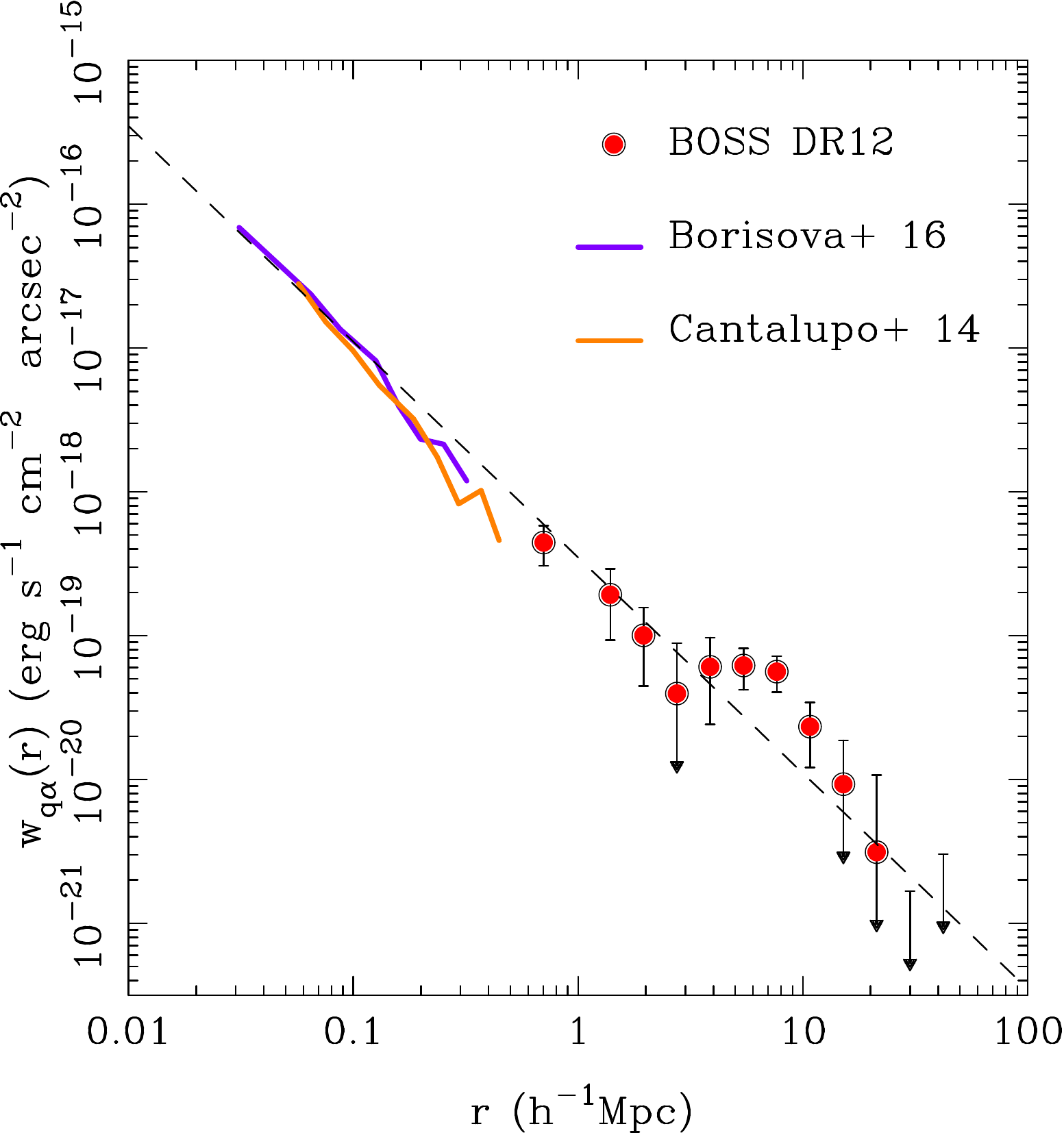}
}
\caption{ The projected quasar-\lya emission cross-correlation
function from BOSS compared to the small-scale results of
Cantalupo et al. (2014) and Borisova et al. (2016). The BOSS results have
been evaluated by projection into a pseudo narrow band with the 
same mean width as used by Borisova et al. (2016)  (see text).
The dashed line represents a power law 
$w_{q\alpha}=3.5\times10^{-19} r^{-1.5} \ergs \cm^{-2} \asec^{-2}$.
\label{proj}}
\end{figure}

In Figure \ref{proj}, we show the Borisova mean \lya profile, as well
as the  circularly averaged \lya surface brightness profile
of the ``Slug'' nebula (Cantalupo et al. 2014). We have used $(1+z)^{4}$
surface brightness dimming to convert the Cantalupo et al. and Borisova et al.  results  to what they would be
at the mean redshift of our present measurement ($z=2.5$).
 In order to compare
to our measurement of \xiqe  we  project our
results to account for the fact that the IFU observations have been 
projected into a pseudo narrow band. Borisova et al. fix the width of their 
pseudo-NB images to the maximum spectral width of the nebulae. These vary
from 105 \AA\  to 23.75 \AA, with a mean width of 43.02 \AA. This mean
filter width  translates to a comoving line of sight distance of 
$r_{\rm NB}=20.75 \hmpc$. We therefore take the \xiqe$(r_{\parallel},r_{\perp})$
measurement shown in Figure \ref{sigpi} and collapse the region
between $r_{\parallel}=\pm20.75/2$ $\hmpc$ along the $r_{\parallel}$ axis
and plot the result as a function of $r_{\perp}$. We label
the result $w_{qe}$, and show it in Figure \ref{proj},
along with the Cantalupo et al. and Borisova et al. results.

We can see from Figure \ref{proj} that although there is not coincidence
in spatial scales between 
the SDSS and IFU meaasurements, the 
scales are almost overlapping. The largest scale datapoint for the IFU data is
at $0.4 \hmpc$, and the smallest SDSS point is at $0.7 \hmpc$. An
extrapolation of the small scale IFU data appears to be reasonably consistent
with the SDSS $w_{qe}(r)$. In order to guide the eye, we plot a power law
$w(qe)=3.5\times10^{-19} r^{-1.5} \ergs \cm^{-2} \asec^{-2}$.
 This curve has no 
physical significance but appears to follow the broad trend seen in the
data. This power law  in \lya surface brightness, seen
over three orders of magnitude in spatial scale may indicate
that the physical process or at least the quasar energetic output
 responsible for the small scale emission profile seen continues to act
detectably 10 $\hmpc$ from quasars.

\section{Lyman-alpha forest-Lyman-alpha emission cross-correlation}
\label{secxife}

If the \lya\ emission seen in Section \ref{secxiqe} were uniformly tracing the
large-scale structure of the Universe, one would expect there to be
significant \lya\ surface brightness in regions that are far from quasars.
The quasar-\lya\ cross-correlation function is not the best way to 
probe this, due to the fact that  the \xiqe measurement is below the noise
level at large quasar-pixel separations. In Figure \ref{cdmfid} we can see
that this occurs at scales $r \ge 15 \hmpc$. The luminosity 
function of SDSS
quasars at redshift $z\simeq2.5$ is
$\Phi=10^{-5.8} {\rm Mpc}^{-3} {\rm mag}^{-1}$ at the $i=21.8$ limit of
the survey (DR9, Ross et al. 2013). The mean interquasar
separation is approximately $50\hmpc$. The volume fraction of space sampled
by quasars is therefore $(15/50)^3=0.03$, and  should be supplemented
by a more space-filling tracer of \lya\ emission to truly test  whether
the \lya\ emission seen in Figure \ref{cdmfid} is due to star forming
 galaxies. 

 The \lya\ forest of absorption by neutral hydrogen in 
quasar spectra offers an alternative. 
The \lya\ forest has long been used as a probe of the cosmic
density field at the relevant redshifts. The physical processes governing
the state of the IGM are simple, and its absorption properties are those
first described by Gunn and Peterson (1965), leading to its characterization as
the ``Fluctuating Gunn-Peterson Effect'' (Weinberg {\it et al.} 1997).
On scales  larger than a pressure smoothing scale (of order $0.1 \hmpc$,
Peeples
et al. 2010),
the forest acts as a biased tracer of the density field.
When dealing with  \lya\ forest clustering it is 
customary to define the ``flux overdensity'', $\df$,
from the observed flux ``F'' in a spectrum as follows:
\begin{equation}
\df=\frac{F}{\langle F \rangle} -1.
\end{equation}
$\df$ is a quantity with zero mean.
On large scales, the quantity $\df$ is  related to the mean
overdensity of matter by  linear bias factor $b_{f}$ Because the \lya\ forest
is saturated in regions of high density, the clustering of the forest
has a relatively low amplitude, and therefore a low bias factor.
McDonald (2003) carried out a determination of the bias factor expected in 
CDM models, finding $b=-0.1511$. 
    This is approximately
consistent with e.g., the measurements of
the \lya\ forest autocorrelation function by Slosar \etal (2011), which yield $b=-0.2\pm0.02$.
Because the quasar flux in a \lya\ forest spectrum is absorbed more 
(lower flux)  in regions of low density and is absorbed less (higher 
flux) in regions of high density, the relationship between $\df$ and 
$\delta$, the matter overdensity has a  negative bias factor, $b_{f}$.
This can be seen in various contexts, such as the negative
cross-correlation function of quasars and the \lya forest (Font-Ribera
et al. 2014). The
amplitude of the \lya\ forest-emission correlation is therefore
expected to be negative in models where high overdensities of matter  (and \lya\ emission)
lead to increased \lya\ absorption.

The forest
has been used in a variety of cosmological measurements, including the
determination of the Baryon Acoustic Oscillation scale from
\lya\ forest clustering at high 
redshifts (Busca et al. 2013, Slosar et al. 2013).
 In our case we will use it to probe cosmic 
\lya\ emission using cross-correlation.
In SDSS DR12, the number
of high redshift $z> 2.15$ quasar spectra is 
175244 over 9376 square degrees of spectroscopic effective area.
This leads to a mean sightline separation of 
comoving $\sim 17 \hmpc$. 
This relatively high density of sightlines makes it
possible to reconstruct the large-scale structure of the cosmic density
field at these redshifts with higher resolution than is possible with
current galaxy or quasar surveys (e.g., Ozbek and Croft 2016).
The cross-correlation function of the \lya\ forest and \lya\ emission,
\xife
will therefore also be much better sampled, with many more \lya\ forest-
\lya\ emission pixel pairs at any separation than was the case
with \xiqe . 

We compute \xife from our data samples in a similar fashion to
the quasar-\lya\ emission cross correlation (Equation \ref{xieq}).
Our estimator is 

\begin{equation}
\xi_{\rm fe}(r) = \frac{1}{\sum_{i=1}^{N(r)}w_{ri}}    \sum^{N(r)}_{i=1} w_{ri}  \Delta_{\mu} \df                  ,
\label{xife}           
\end{equation}
where $\delta_{\rm F}$ is the \lya\ forest flux overdensity.

We use the same 160 subvolumes of the survey to construct jackknife
samples, and use these to compute errors bars as was done with \xiqe.
As with \xiqe, we have found that there is some cross-fibre light which 
could affect the measurement. We again do not use pairs of \lya\ forest and 
\lya\ emission pixels which are separated by 5 fibres or less in 
computing Equation \ref{xife}. After doing this, again as with \xiqe,  a
 small amount of residual light contamination remains due to
quasar clustering. This can be removed either by subtracting a model
for the contamination or by completely removing all fibres with $\deltafibre
\leq 5$ from the sample. In Appendix A we  carry out tests on both of these 
methods, and show that there is no significant difference in our conclusions
when either is used, or even if the contamination is not corrected
for (it is very small in the case of \xife).

\begin{figure}
\centerline{
\includegraphics[width=0.45\textwidth]{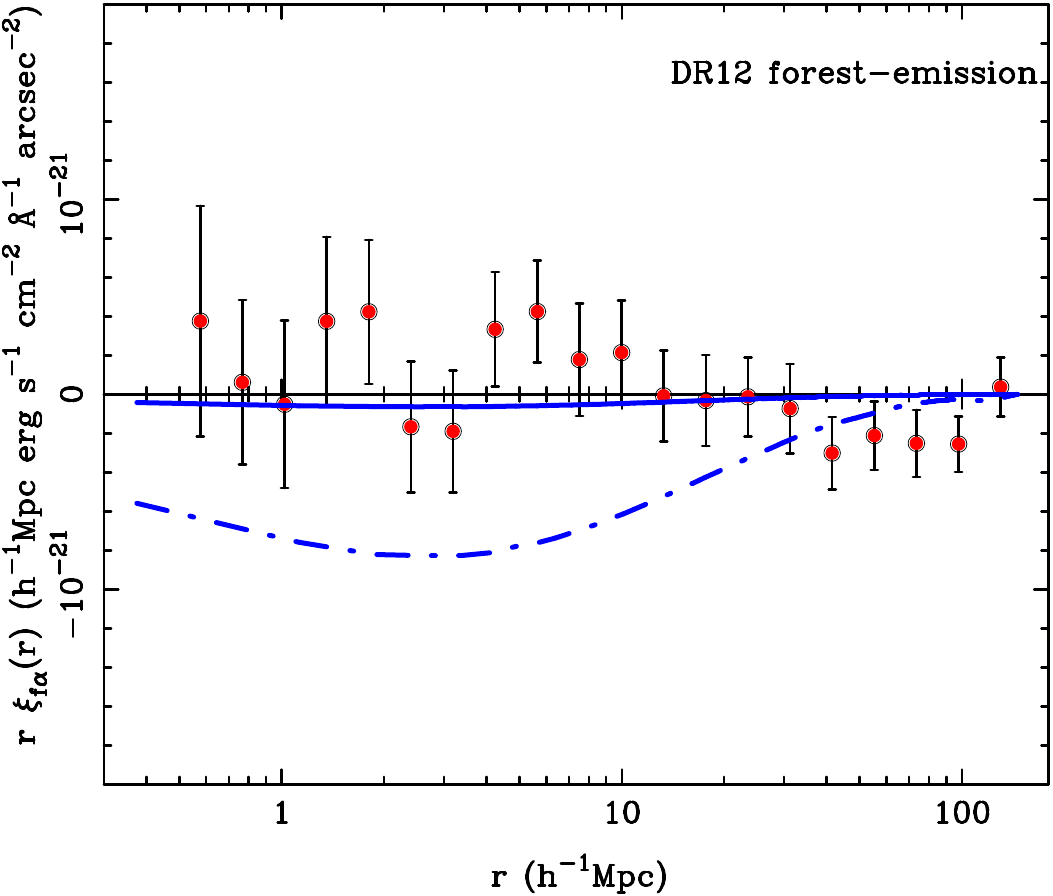}
}
\caption{ The cross-correlation function of \lya emission and the \lya
forest, $\xi_{f\alpha}$ for spectra from  BOSS DR12 (points with error bars)
as a function of \lya emission - \lya forest pixel pair separation.
 The solid
line shows the  predicted cross-correlation function 
if the \lya emission were tracing the large-scale structure of
the Universe (for example being caused by star forming galaxies), and 
the mean surface brightness of \lya\ emission in the
Universe is given by the contribution of all individually
detected \lya\ emitters. 
It 
should be noted that the predicted amplitude of  $\xi_{f\alpha}$
is negative (because the \lya forest has negative bias). The dashed line shows the predicted cross-correlation function                       
if the \lya emission were tracing the large-scale structure of the Universe but
the mean surface brightness of \lya\ emission was at the very high level
needed to account for the results in G.
This is clearly ruled out, indicating
that the mean surface brightness of \lya\ emission must be at a lower level.
\label{fexi}}
\end{figure}

 In Figure \ref{fexi} we show
our results (in this case the modelled contamination has been subtracted). We
can see that there does not appear to be any strong evidence for a
non-zero \xife\ signal. We will see later that a model fit shows that this
is indeed a null result.  
Comparing Figure \ref{fexi} to Figure \ref{cdmfid}, the y-axis scale has been
magnified by a factor of 10, so that the overall signal in the quasar-emission
correlation would be completely off the top of the panel in the current
plot. Because the bias factor of the forest is much lower, however, one would 
expect the \xife\ signal to be much smaller than \xiqe. We now examine this 
expectation in the context of the model where the \lya\ emission surface 
brightness traces the large-scale structure of the Universe.

We have seen in Section \ref{modelfit} that if the \xiqe\ signal seen is due to
star forming galaxies which trace structure, then a very high mean
\lya\ surface brightness of 
$\langle \mu_{\alpha} \rangle=(1.9 \pm 0.5) \times10^{-21}                       (3/b_{\alpha}) \ergs\cm^{-2} \angs^{-1}\asec^{-2} ~.$                             is inferred, and 
this \lya\ emission is associated with a star formation rate 
${\rho}_{\rm SFR}(z=2.55) = (0.14\pm0.04) (3/b_{\alpha})                   \msun\, {\rm yr}^{-1} \mpc^{-3} ~.                    $                     
In C16, a qualitatively similar conclusion was reached (although the 
results were approximately a factor of two higher due to the presence of 
contamination from quasar clustering). We are now in a position to 
test this model, as it predicts that for \xife\ we should see the
same shape as \xiqe from Figure \ref{cdmfid}, but with the amplitude
scaled down by a factor of (-0.3/3.6), which is the
ratio of the \lya\ forest bias factor to the quasar bias factor. This value of -0.3 for the forest bias factor is approximate (see Slosar et al 2011), and includes the effect of redshift space distortions ($bf_{\beta}=-0.3$, see Section 4.1). Quasar redshift distortions have a negligible effect on the clustering amplitude in this context.
We have plotted this prediction as a dot-dashed line in Figure \ref{fexi}.
We can immediately see that it is not consistent with the DR12 results,
which indicates that the \lya\ emission seen in \xiqe\ cannot
be spread throughout space with a high surface brightness.

The other solid line in Figure \ref{fexi} shows the predicted \xife\ curve
that corresponds to the same model, but with a much lower mean
surface brightness of \lya\ emission, that due to the summed
emission of known \lya\ emitters. The results of Gronwall et al. (2007)
 have shown that
these correspond to a star formation rate at $z\sim2.5$ observed
through \lya\ of ${\rho}_{\rm SFR}=0.01 \msun {\rm yr}^{-1}$. This is a factor of $\sim15$
 smaller than the 
high surface brightness model. By eye, it is apparent that this very low
amplitude curve is not very different from zero given the error bars
of the DR12 result. As such, the observed \lya forest-\lya emission
cross-correlation appears to be consistent with  known \lya\ emitters.
It is however possible to use \xife to place limits on 
the presence of other \lya\ emission
that traces cosmic structure, including very low surface brightness
emission that would not have been detected in \lya\ emitter surveys.

\begin{figure}
\centerline{
\includegraphics[width=0.45\textwidth]{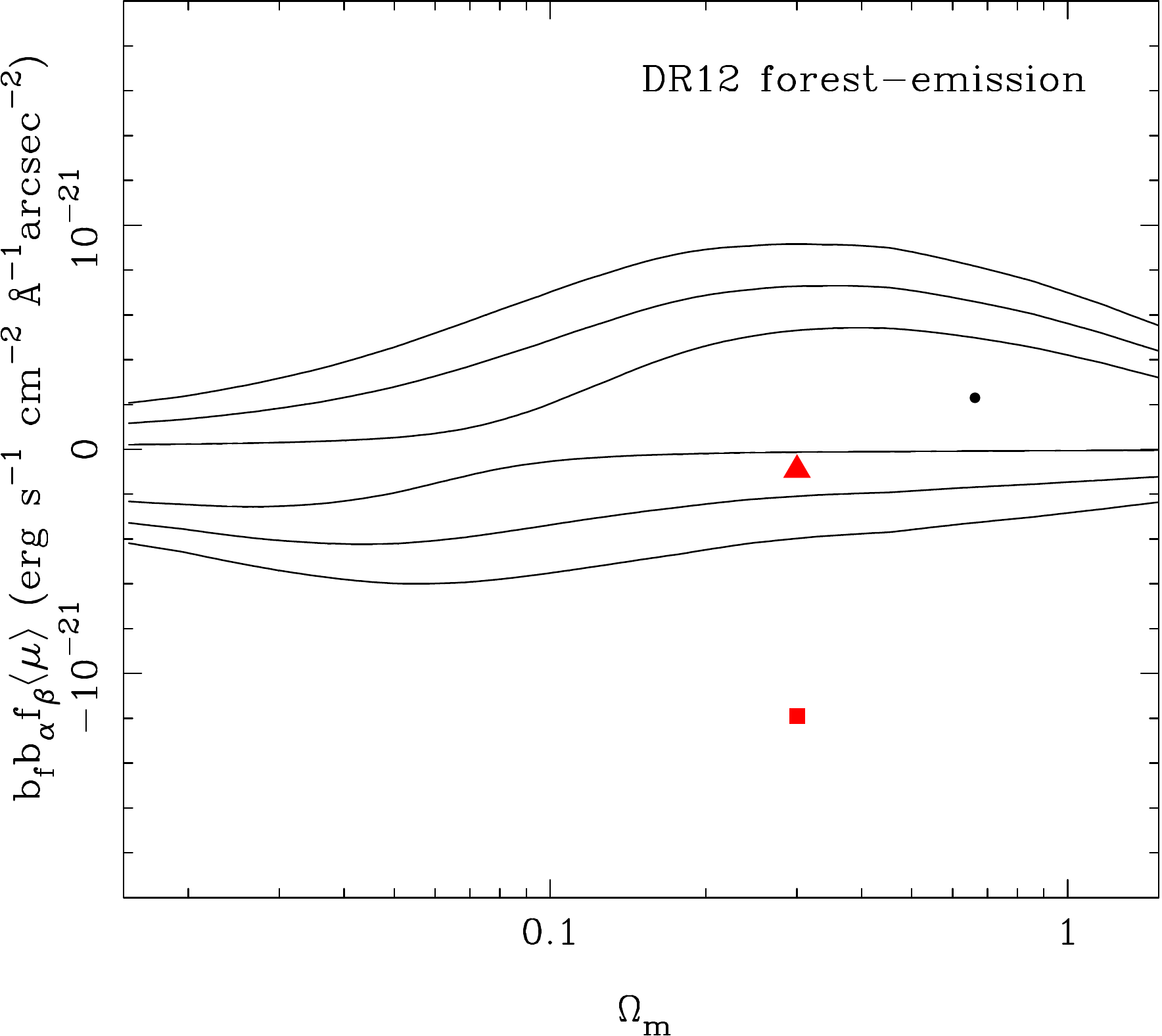}
}
\caption{
Fit parameters for the amplitude
\ampfa\  and shape $\Omega_{\rm m}$ (for fixed $h$ and other parameters)
of a linearly biased CDM model fit to the forest-\lya emission
 cross-correlation
function plotted in Figure \ref{fexi}. The dot indicates the best fit
parameters and the contours show the 1, 2 and 3 $\sigma$ confidence contours.
The triangle corresponds to the parameters for the solid theory line
in Figure \ref{fexi} (mean \lya\ emission consistent with known \lya\ emitters)
and the square the dot-dashed theory line in Figure \ref{fexi} ( mean
\lya\ emission consistent with emission tracing large-scale structure,
model G 
and an amplitude needed to fit the \xiqe observations in this model).
 This model is
clearly ruled out.
\label{fega}}
\end{figure}

\subsection{Linear CDM fit to forest-emission cross-correlation: model G}

We do this by carrying out model fitting, using the same biased
linear CDM correlation function used in Section \ref{modelfit} (model G).
 The amplitude parameter
in  the present case is \ampfa\ , and the shape parameter is
again $\Omega_{\rm m}$. In figure \ref{fega} we show  the contours 
of $\Delta\chi^{2}$ in this parameter space. We can see that the
best fit model has a positive amplitude (the opposite
sign to that expected for \xife), but that it is consistent
with zero at the $\sim 1 \sigma$ level, as we expected given our
visual impression of Figure \ref{fexi}. The best fit parameters
are
\begin{equation}                                                               
\text{\ampfa}  = (2.5\pm{1.8}) \times 10^{-22} \ergs \cm^{-2} \angs^{-1}
\asec^{-2},                       
                          \end{equation}
and $\Omega_{m}=0.691^{+2.06}_{-0.47}$.
In Figure \ref{fega} we have plotted
symbols representing the high surface brightness \lya\ model,
and the \lya\ model representing known \lya emitters. The former lies at
a $\Delta\chi^2=56.5$ from the best fit, indicating that it is ruled out
at the $7.5 \sigma$ level. The latter is within $\Delta\chi^2=3.7$ of
the best fit, indicating that it is acceptable at the $1.9 \sigma$
level. Assuming a fixed shape parameter of $\Omega_{\rm m}=0.3$
leads to a 95 per cent lower limit on the
 parameter \ampfa$=-1.07\times10^{-22} \ergs \cm^{-2} \angs^{-1} \asec^{-2}$
The 95 per cent upper limit on the mean \lya surface brightness is
then 
\begin{equation}                                                                
\langle \mu_{\alpha} \rangle < 1.2 \times10^{-22} 
 \frac{-0.3}{b_{f}f_{\beta}}     
(3/b_{\alpha}) \ergs\cm^{-2} \angs^{-1}\asec^{-2} ~.                        
\label{eq:muf}                                                                
\end{equation}
Here we assume fiducial values of 
$b_{f}f_{\beta}=-0.3$ (motivated by Slosar et al. 2011) and $b_{\alpha}=3$. This
limit is in the context of model $G$, and is  a factor 
of 15 lower than the value estimated from \xiqe in Equation \ref{eq:backe}.
The 95 per cent upper limit on the associated star formation rate density is
therefore also a factor of 15 lower, and just consistent with the measurement
from known \lya emitters. 

\begin{figure}
\centerline{
\includegraphics[width=0.45\textwidth]{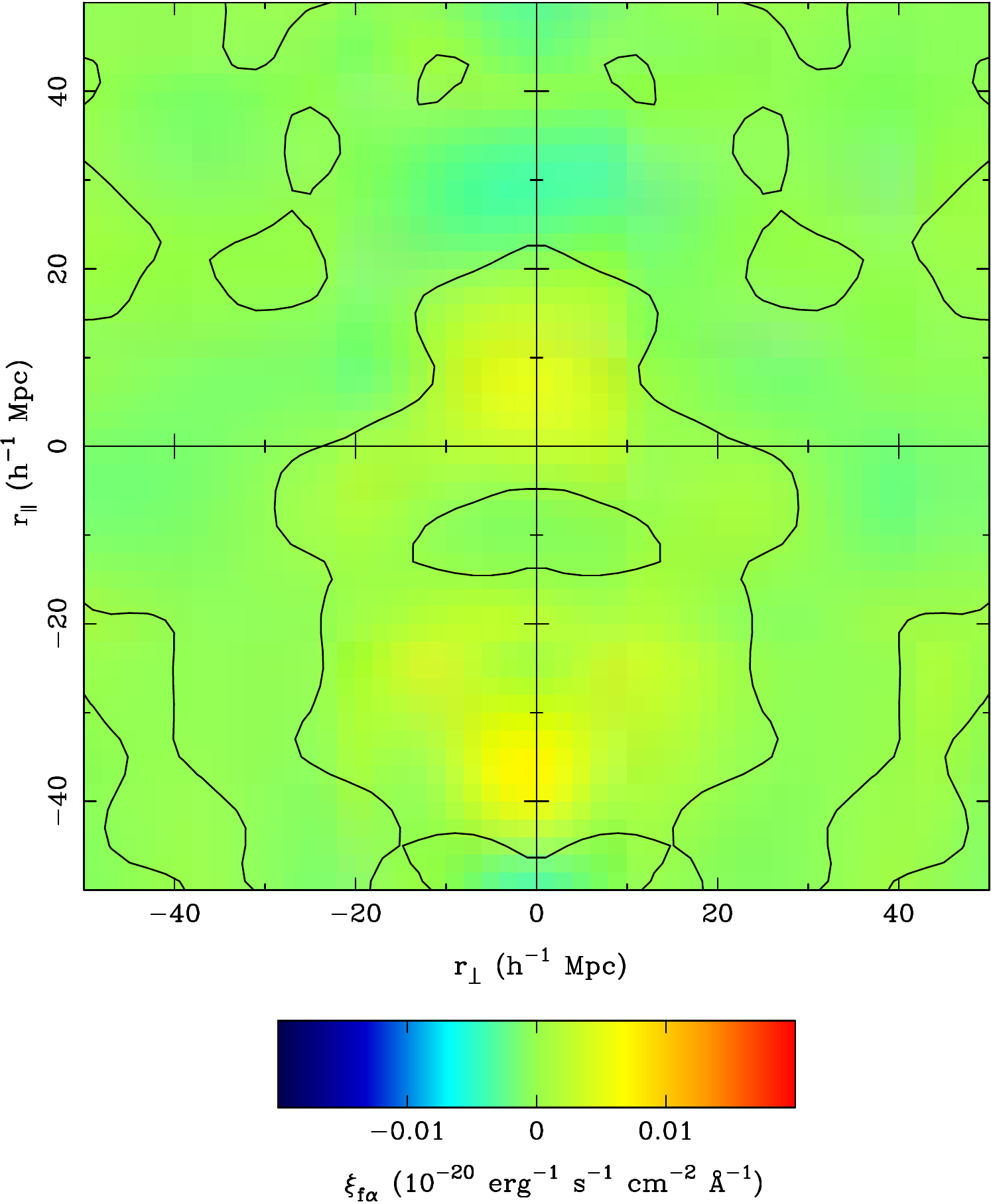}
}
\caption{ The cross-correlation function of \lya emission and the \lya
forest, $\xi_{f\alpha}$ for spectra from  BOSS DR12 (points with error bars)
as a function of separation parallel to and perpendicular to the line 
of sight.  The predicted amplitude of  $\xi_{f\alpha}$
is negative (because the \lya forest has negative bias). The cross-correlation
function plotted here is consistent with noise.
\label{fexisp}}
\end{figure}

In order to confirm this null result, 
we can also examine \xife as a function of pixel separation 
across and along the line of sight. This is shown in Figure \ref{fexisp}.
The expected signal in the event of significant \lya\ forest- emission
cross correlation would be negative, i.e. on the blue end of the colour table.
We can see that instead the plot is mostly green, indicating no signal. Any
hint of a positive correlation in the radially averaged version of
this plot (Figure \ref{cdmfid}) is seen here to result from a faint blob
which  is off center, this fact, and the $\sim 1 \sigma$ signficance
both point to their being an absence of signal.

There is therefore little room for excess surface brightness over that
contributed by known \lya\ emitters in this model of \lya\ emission
tracing large-scale structure. The question still remains whether there
is a way that \lya\ emission can be spatially distributed in a fashion 
which is consistent with both the \xiqe and \xife  constraints. 
In order to address this we have carried out some simple theoretical
modelling using a cosmological hydrodynamic simulation, as described
in the next section.

\section{Comparison to simulations}

\label{secsim}

The \lya surface brightness seen within 15 $\hmpc$ of quasars is certainly
substantial (Figure \ref{cdmfid}). We would like to  
know whether it is possible
to model  such a  high surface brightness and yet not breach the 
\xife constraint
on \lya\ emission from the more widely distributed intergalactic medium. 
To answer this, we 
set up some simple toy models using a cosmological simulation as a base. 

\subsection{Simulation model}

\label{secsim}

In order to resolve the relevant
pressure smoothing scale in the forest and
large-scale structure, we use a large hydrodynamic
cosmological simulation of the $\Lambda$CDM model. The smoothed
particle hydrodynamics code {\small P--GADGET} (see
Springel 2005, Di Matteo et al. 2012) was used to evolve
 $2 \times 4096^2 = 137$ billion particles in a cubical periodic 
volume of side length 400 $\hmpc$. This simulation was previously
used in Cisewski et al. (2014) and Ozbek and Croft (2016), where more details
are given.

 The simulation
cosmological parameters were $h = 0.702,\, \Omega_{\Lambda} = 0.7
25,\, \Omega_m = 0.275, \, \Omega_b = 0.046, \, n_s = 0.968\; $and$ \;
\sigma_8 = 0.82.$ The mass per particle was $1.19 \times 10^7$ $h^{-1}
M_{\odot}$ (gas) and $5.92 \times 10^7$ $h^{-1} M_{\odot}$ (dark
matter).  An ultraviolet background radiation field consistent with
Haardt and Madau (1995) is included, as well
as cooling and star formation. The latter, however
 uses a lower density threshold than usual
(for example in Springel \& Hernquist 2003) so that gas
particles are rapidly converted to collisionless gas particles.  This
is done to speed up execution of the simulation. As a result the
stellar properties of galaxies in the simulation are not predicted
reliably but this has no significant effect on the diffuse IGM that
gives rise to the Lyman-$\alpha$ forest. We do not otherwise use the galaxies
in our modeling, but instead use the overall baryonic density
field to generate a biased \lya\ emission spatial distribution (see below).

\begin{figure*}
\centerline{
\includegraphics[width=0.9\textwidth]{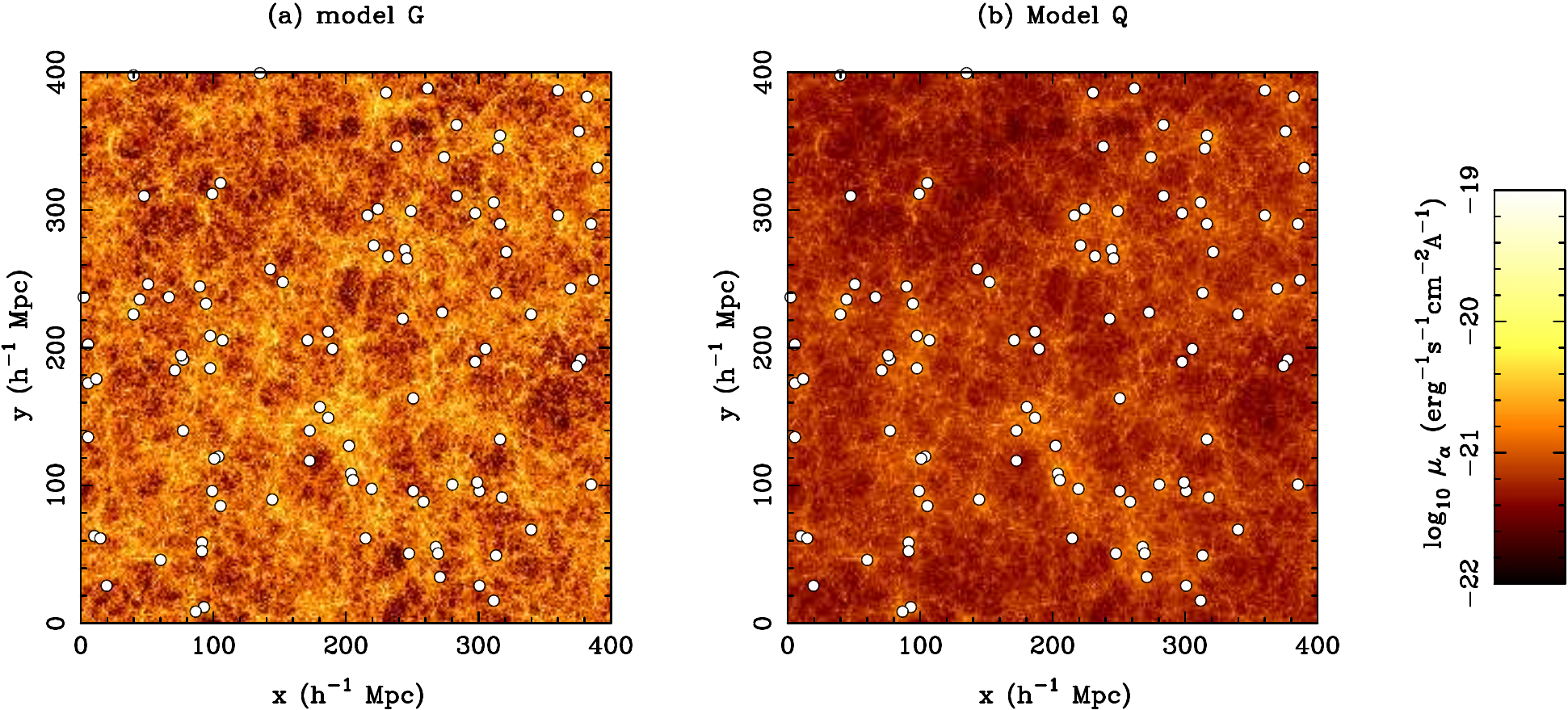}
}
\caption{Slices ($40 \hmpc=10\%$ of the box thick)
through the simulation \lya emission field in the 
two toy models model of Section \ref{secsim}. The \lya emission
surface brightness is shown by a colour scale and the positions of
bright quasars by points. The quasar-\lya emission cross-correlation
function in both cases is consistent with the SDSS observations in 
this paper, but model G overpredicts the \lya forest-emission cross-correlation
by a factor of $>10$.
  \label{modelslice}}
\end{figure*}

 The simulation snapshot at redshift $z = 2.5$ 
(the mean redshift of our SDSS observations) is used
to generate a set of Lyman-$\alpha$ spectra using information
from the particle distribution (Hernquist et al. 1996).
The spectra are generated on a grid with $256^2 = 65536$ 
evenly spaced sightlines, resulting in 1.56 $\hmpc$ spacing.
We also use the simulation particles to generate
a baryonic matter density field sampled along the same sightlines. This field
will be used to model the \lya\ emission. 

We also use the baryonic matter
density to generate quasars. To do this, we resample each sightline so that
it has 256 pixels (the density field is now a $256^3$ grid). We find all
the local maxima in that grid and select the 512 with the highest
local density to be the
locations of quasars in our model. The mean quasar separation is therefore
$50 \hmpc$, approximately consistent with the mean separation in the BOSS 
sample
of quasars (Ross et al. 2013). 
In order to to further check their suitability 
 we compute the autocorrelation function of the simulated quasars, $\xi_{qq}$.
We find that $\xi_{qq}$, while noisy as expected with not many objects, it
has a shape that is approximately consistent with the CDM autocorrelation
function appropriate for the simulation.
 We assign 
Poisson errors to the quasar autocorrelation function data points 
and carry out a fit of the bias factor, $b_{q}$ relating the linear matter and 
quasar autocorrelation function over scales from $r=4 \hmpc$ to $r=70 \hmpc$.
We find $b_{q}=4.1 \pm 0.8$, consistent at the  $1 \sigma$ level with the
$b_{q}=3.6$ (Font-Ribera et al. 2013) from the SDSS/BOSS quasar sample.

The simulation \lya\ emission in the model is derived from the baryonic
matter density field. We try two variations of the model:\\
\\
Model G. The \lya\ emission intensity is directly proportional
to the matter density. This model represents \lya\ emission uniformly tracing
the large scale structure of the Universe (albeit in a biased
fashion), for example being due to 
star forming galaxies. We have
\begin{equation}
\mu_{{\rm ly}\alpha}({\bf  \rm x}) = \langle \mu \rangle b_\alpha \rho({\bf \rm x}),
\end{equation}
where $\rho({\bf \rm x})$ is the baryonic matter density in units of the mean at 
a point separated by vector ${\bf \rm x}$ from the coordinate origin.\\
\\
Model Q. We assume that quasars are responsible for \lya\ emission intensity
with a \lya\ surface brightness that is proportional to 
the product of the density and  the inverse of the square of the
 distance from a quasar.
 We have for a given \lya\ emission pixel,
\begin{equation}
\mu_{{\rm ly}\alpha}({\bf \rm  x}) =\mu_{0} \rho \Gamma ({\bf \rm x}),
\label{eqmu}
\end{equation}
where $\Gamma ({\bf \rm x})$ is the radiation intensity field 
computed in the optically thin limit,
 assuming all quasars have the same luminosity, $L_{0}$,
and $r_{i}$ is the distance from the point in question to quasar $i$: 
\begin{equation}
\Gamma({\bf \rm x}) =\sum_{i=1}^{n_q} L_{0} \frac{1}{r_{i}^{2}} e^{-r_{s}/r_{i}}.
\label{eqgamma}
\end{equation}
Here the product $L_{0}\mu_{0}$ is a 
constant, a free parameter which we set by fitting the amplitude
of \xiqe in the model to the observational data from SDSS/BOSS DR12.
This model is meant to approximate scenarios where the energy from 
quasars is distributed to nearby gas following an inverse square law
and induces \lya\ emission. This could be through fluorescent emission
e.g., Kollemeier et al. (2010),
hard quasar ionizing radiation heating the gas, or other mechanisms. 
The  $e^{(-r/r_{s})}$ term smooths the \lya intensity field on small 
scales (we use $r_{s}=2 \hmpc$). Without it, the small scale behaviour
of \xiqe\ is too steep. We leave investigation of the physics of
the region within 2 $\hmpc$ of quasars to more sophisticated 
future simulations
than our toy models.

In Figure \ref{modelslice}
 we show slices through the \lya\ emission field in our
two models. In each case, we have adjusted the mean \lya\ surface 
brightness by tuning the parameters product
$L_{0}\mu_{0}$ in Equations \ref{eqmu} and \ref{eqgamma} so 
that the quasar-\lya emission cross-correlation function in the  models 
is consistent with the results from SDSS DR12 (see next section).
We can see that in the Model G case  the emission does indeed trace
structure, and prominent filaments can be seen. Looking at the Model Q panel,
there is much more inhomogeneity in the emission. By design, the 
\lya\ surface brightness close to quasars is similar to Model G, but
in the regions far from quasars there are darker voids in 
the \lya\ emission. The mean \lya\ surface brightness is also 
lower in Model Q. We find 
$\langle \mu_{\alpha} \rangle = 1.5 \times10^{-21}  
\ergs\cm^{-2} \angs^{-1}\asec^{-2}$ for model G (assuming $b_{\alpha}=3$),
which is consistent with what was found from the linear theory
model G in Section \ref{modelfit}. For model Q, we 
find   $\langle \mu_{\alpha} \rangle = 7.0 \times10^{-22}       
          \ergs\cm^{-2} \angs^{-1}\asec^{-2}$. Although the 
$\langle \mu_{\alpha}\rangle$ values are only a factor of $\sim 4$ apart for
model G and Q, we see below that the bias factor $b_{\alpha}=3$ 
applicable only to model G strongly affects the clustering measures
\xiqe\ and \xiqe\ .

\subsection{Quasar-Lyman-alpha emission cross-correlation in simulations}

\begin{figure}
\centerline{
\includegraphics[width=0.45\textwidth]{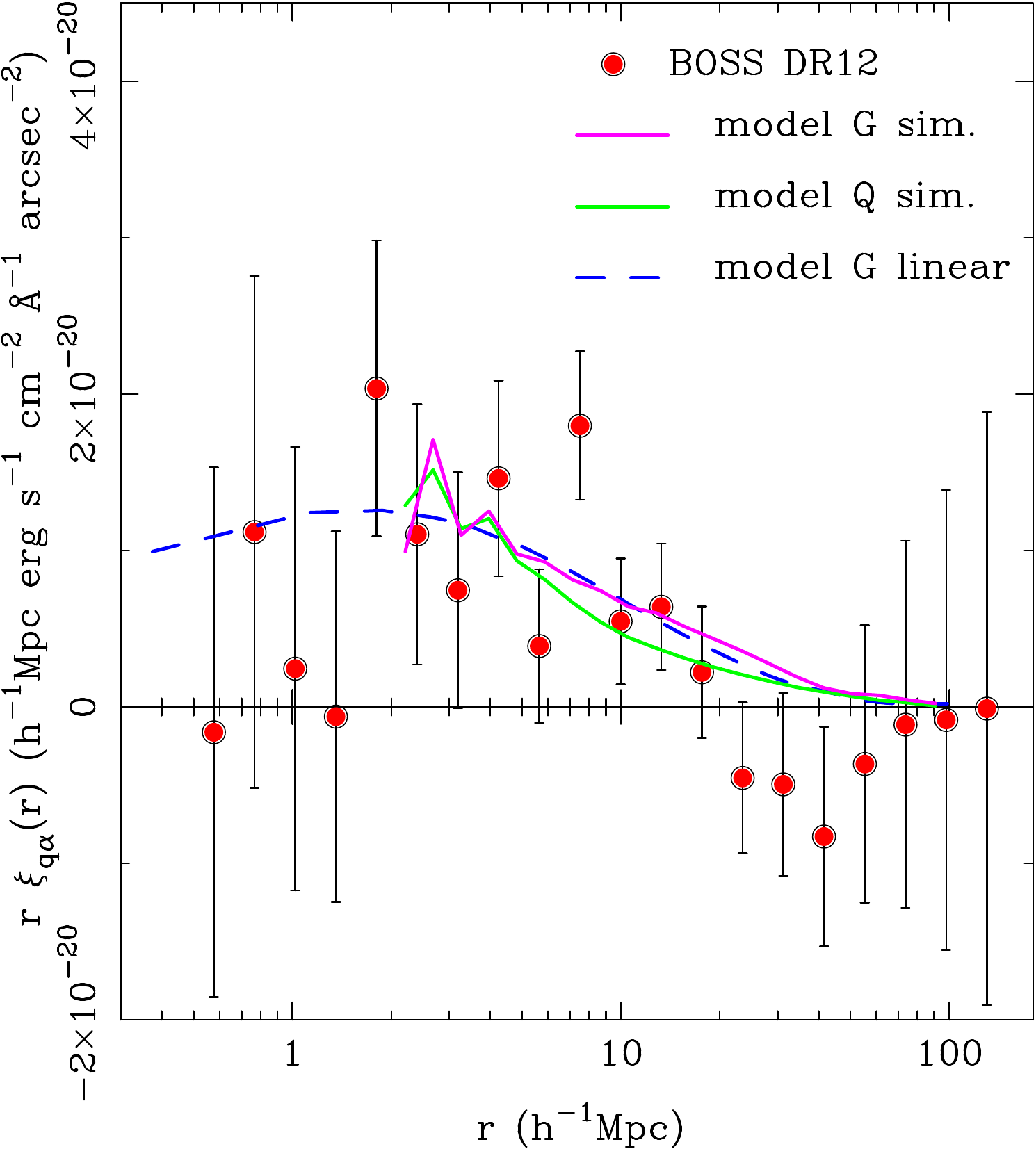}
}
\caption{ The quasar-lya emission cross-correlation function,
in the simulation-based models G and Q, as well as the linear theory
version of model G. The SDSS/BOSS data points are also shown.
\label{modelqe}}
\end{figure}

We next compute \xiqe for the quasars and the model G and model Q \lya\ 
emission fields in the  simulation.
 First we move the quasars and emission field into redshift space.
We use one axis of the simulation
volume as the line of sight, moving quasars to their redshift space positions
and convolving the \lya\ emission field with the line-of-sight peculiar
velocity field.
 We apply Equation \ref{xieq} to the resulting quasar distribution
and emission field, giving \xiqe results which are shown in 
Figure \ref{modelqe}. Because of the limited resolution of the simulation
(we are working with sightlines spaced by $400/256=1.6 \hmpc$), there is
no clustering information below $r= 2 \hmpc$.

We can see that \xiqe for model G is approximately consistent with 
the linear theory model which was earlier (Section \ref{modelfit})
 found to be a reasonable
fit to the observational data. This is expected, the simulation model 
was designed to be qualitatively the same, i.e. \lya\ emission being a biased
tracer of the mass distribution.

Looking at the results for model Q, we can see
that the amplitude of \xiqe 
 is also similar to the observations (given the large error bars)
and to model G. The lack of emission seen in the void regions
in  Figure \ref{modelslice} does not affect the fit, showing once more
that only the emission with $\sim 15 \hmpc$ of quasars is relevant to \xiqe.

\begin{figure}
\centerline{
\includegraphics[width=0.45\textwidth]{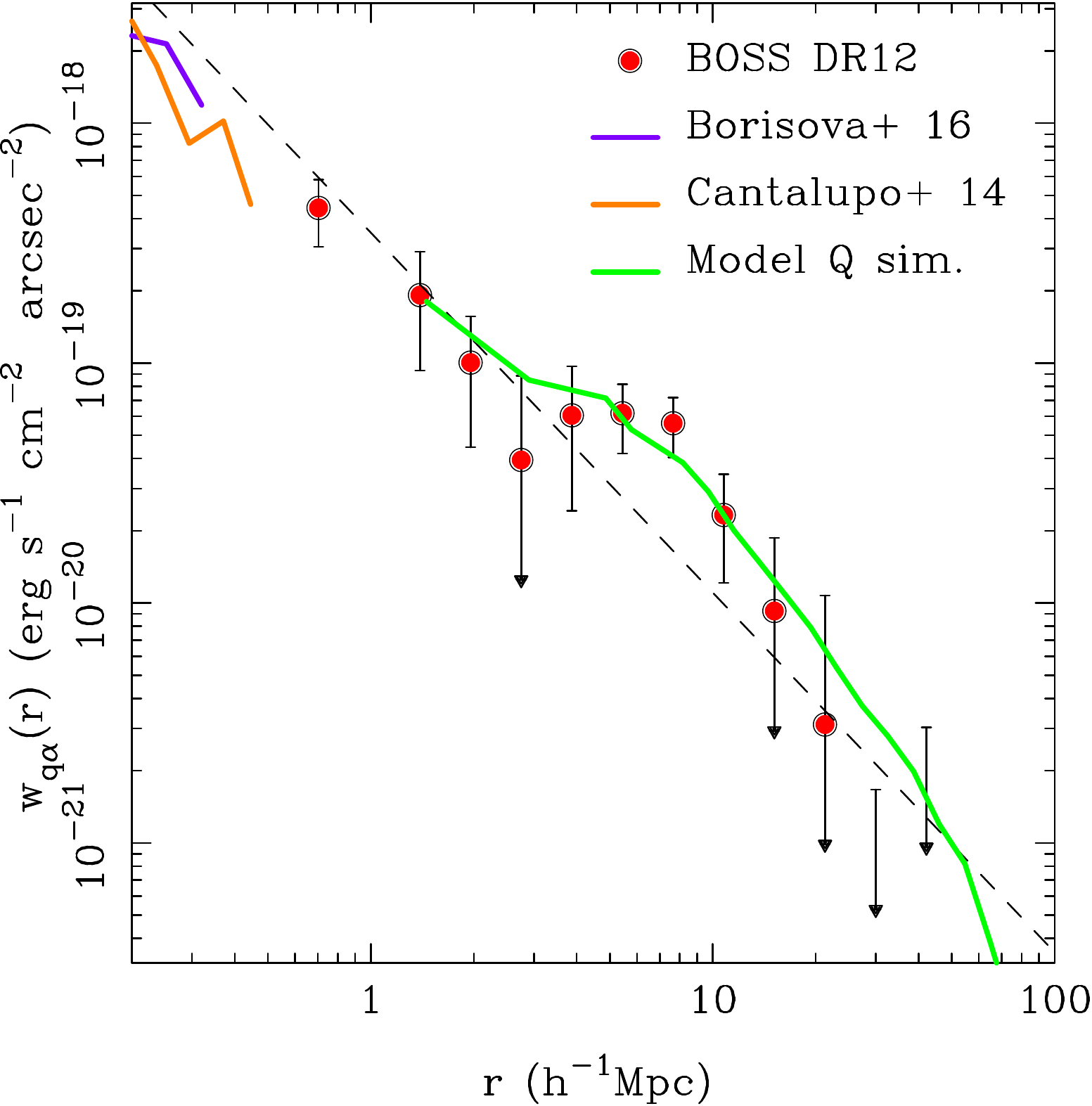}
}
\caption{ The projected quasar-lya emission cross-correlation function
in the simulation of model G, together with the observational data points from 
Figure \ref{proj}.
\label{projsim}}
\end{figure}

In Figure \ref{projsim}, we show the projected quasar--lya emission 
cross-correlation function, \wqe for model Q, along with the
SDSS data points and the larger scale of the quasar \lya blob data 
from Figure \ref{proj}. The simulation and observations are in 
reasonable agreement.

\subsection{Forest-Lyman-alpha emission cross-correlation in simulations}

\begin{figure}
\centerline{
\includegraphics[width=0.45\textwidth]{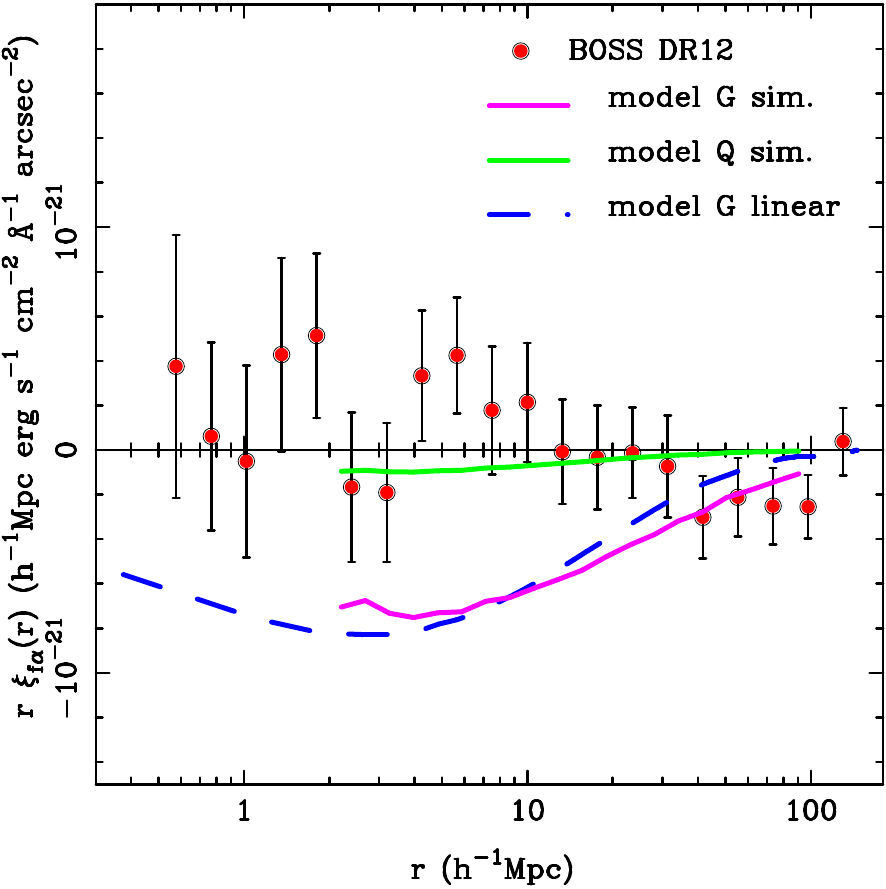}
}
\caption{ The forest- \lya emission cross-correlation function in simulations (solid lines, purple for model G and green for model Q) and the SDSS DR12 observations (points with error bars). The dashed line shows the results for the linear theory version of model G.
  \label{modelfe}}
\end{figure}

We use the same simulations  to compute \xife, again with the
estimator of equation \ref{xife}. The 
results are shown in Figure \ref{modelfe},
where it can be seen that the  Model G simulation 
curve follows the curve for the linear
theory model G, although not
as closely as for \xiqe (Section \ref{modelfit}).
 Both it and the linear theory curve
are significantly larger in amplitude
(while being negative in sign as expected because of the 
negative bias of the forest) than the observations. Model Q, on the other hand
is much smaller in amplitude, reflecting the fact that most of the volume
of space is far from quasars and therefore has little \lya\ emission. 
The model Q results for \xiqe and \xife show that it is
possible to realize a large-scale distribution of \lya surface brightness which
is consistent with both sets of observations. This was not trivial, as it could
have been the case that both observations were mutually inconsistent, which
would have meant that there was some problem with the measurement.

While we have shown that it is possible to distribute \lya surface brightness
in a way which matches observations (at least on scales $r > 2\hmpc$), we have
done this using a toy model. In future work that extends that 
of e.g., Kollmeier et al. (2010) and Kakiichi and Dijkstra (2017),
 it will be interesting to see 
if a first principles
physical model is able to reproduce both the Borisova et al (2016) 
results and our observational data on large scales. Radiative transfer will
be important to understand the \lya emission, but it may also be relevant
when considering the \lya forest and \xife. For example, we have not 
included the quasar proximity effect (Bajtlik \etal 1988) in our modelling, which could
suppress \xife.

\section{Summary and Discussion}

\subsection{Summary}
We have searched for the signature of large-scale structure in the
\lya\ emission intensity of the Universe between redshifts
$z=2-3.5$, using cross-correlation techniques
applied to data from SDSS/BOSS DR 12.
Our findings are as follows:

(1) We have detected structure on scales from $1-15 \hmpc$ in the
cross-correlation of quasars and \lya\ emission, \xiqe. 
The shape of \xiqe on these scales is consistent
with the linear ${\rm \Lambda}$CDM shape, as seen in our earlier work (C16). 
Improving on the earlier work, we have identified a  source of
light contamination not previously accounted for, due to the effect
of quasar clustering on cross-fibre light. The amplitude we find
is lower by 50\% than in C16 because of this, but our conclusions with
respect to \xiqe are not qualitatively changed.

(2) We do not detect any signal when cross-correlating the 
\lya forest flux in spectra with the \lya emission samples ( the relevant
statistic is \xife ). 
This non-detection allows us to place limits on the mean surface brightness
of \lya emission in a model where the emission traces the biased matter
density field. The upper limit from this is
$\langle \mu_{\alpha} \rangle < 1.2 \times10^{-22}  \ergs\cm^{-2} \angs^{-1}\asec^{-2}$ at the 95\% confidence level.                      
The corresponding upper limit on the associated star formation rate density 
is 
${\rho}_{\rm SFR}(z=2.55) < 0.01 \left({3\over b_{\alpha} }\right)           
      \msun\, {\rm yr}^{-1} \mpc^{-3} ~.                    $                     
This is the same value as that measured 
from individually detected \lya emitters.

(3) We have used cosmological hydrodynamic simulations to jointly 
examine \xiqe and  \xife in toy models where \lya emission traces
either the large-scale structure in the star forming galaxy distribution
(model G), or is associated more locally with quasars (model Q). In model
Q, we attenuate the \lya surface brightness around 
quasars with an inverse square law, and as a
result the 97\% of the volume of space more than $15 \hmpc$ from a quasar
has a very low level of \lya emission. We find that only model Q can match
the observational measurements of both \xiqe and \xife. 

(4) We have computed the  projected \lya surface brightness profiles
around SDSS quasars by projecting our \xiqe results into
a pseudo narrow band. Extrapolating our results to small scales
 using a power law we
find approximate consistency with the projected \lya profiles measured
on $< 0.5 \hmpc$ scales
from the bright \lya blobs seen around quasars by Cantalupo et al. 2014 
and Borisova \etal 2016.  

(5) Taken together, (1-4) above make it likely that the \lya emission
detected in \xiqe is due to reprocessed energy from the 
quasars themselves and the \lya emission
from star forming galaxies is  at a level not much different with that from
individually detected \lya emitters.

\subsection{Discussion}

In our previous work on the \lya emission-quasar cross-correlation (C16)
we explored possible interpretations for the signal. It was estimated that 
it is extremely unlikely that fluorescent emission due to quasar radiation
is responsible, but that reprocessed HeII ionizing radiation from quasars
or heating from quasar jets are both feasible on energetic grounds. The other
prominent possible explanation  was that the \lya emission seen was
due to escape of \lya radiation from star forming galaxies. Converting
the \lya surface brightness into a mean star formation rate density gave
a surprisingly high value, which was similar to the estimated 
total dust-corrected SFR density, and $\sim 30$ times higher than
the SFR density of known \lya emitters. Because of an additional
correction for light contamination, our present results are lower by
approximately a factor of two, but because of the uncertainty in the
total SFR density this does not change the overall conclusions. In our
current work, we have seen that such a scenario is inconsistent with
another observable, the \lya emission-\lya forest cross-correlation, \xife
and this therefore seems to leave quasar emission as the source of
the \lya emission as the only possible astrophysical explanation.

The non-detection of \xife can be used to place interesting 
limits on the \lya emission from galaxies that trace the large-scale
 structure. We find a 95\% upper limit, which is very close to the value from
 known lya emitters. This means that we should see their effects with
other samples in the near future with this
type of analysis. 
We note that in Equation \ref{eq:muf} where this limit is given,
the limit scales like $3/b_{\alpha}$. This means that if the
luminosity weighted bias factor of \lya emission is lower than 3, the limit
is less restrictive. This is likely if the \lya emission is primarly from 
\lya\ emitters rather than all galaxies (e.g., the bias of measured by
Gawiser et al. 2007 and Gauita et al. 2010 is closer to 1.8).
Nevertheless, the fact that we are close to the individually
detected value means that it is not
possible for the known \lya emitters to be surrounded by very extensive 
halos of low surface brightness \lya emission or very faint \lya emitting 
galaxies, at least not enough to increase the total \lya\ luminosity 
density by a factor $\sim 2-3$. This constraint is close to the amount seen 
in extremely deep imaging of \lya\ emitters by Steidel \etal (2011),
Matsuda \etal (2012).
 and Momose \etal (2014)

We have removed light contamination from our measurements of
\lya intensity clustering, but it is still possible that some
unidentified light contamination still exists. This is a general
problem for intensity mapping approaches to studying large-scale
structure, as well as light from interloper lines (Pullen et al. 2016) the
effect if which is mitigated by cross-correlation. 
Nevertheless, these 
kind of techniques, particularly cross-correlation with multiple
tracers if some are are available
should be useful in eliminating systematic errors. Once 
a detection of \xife is made, and it is consistent with measurements
of other related statistics this would help significantly as a check
on our result. The fact that the \xiqe results from SDSS are quite similar
to an extrapolation of from sub megaparsec scales of quasar \lya blob 
profiles (e.g., Borisova et al 2016) is supportive  of the measurement.

Neverthless, we may be  at the limit of what can be done with this 
kind of non-specialized dataset. We are using a set of fibre spectra which 
were not designed for IM, and 
we are severely limited by light contamination. In future IM work
it should be a priority to design the dataset to minimize light
contamination. Wide field integral field spectroscopy,
such as that being carried out by the HETDEX (Hill \etal 2016) should easily
detect the signal that we have seen, with much 
better control over systematic errors.

We leave theoretical considerations related to the
\lya emission from quasars which is likely responsible for the
signal we have seen to future work. In C16, we saw that it is
energetically possible for quasars to be involved, 
but the detailed mechanisms should be studied using 
physical modelling, such as with hydrodynamic simulations and
radiative transfer. The toy models we have considered in this
paper involve limited actual modelling of physical
processes. There are also several physical effects which have not
been included at all, for example the so called ``proximity effect'' (Bajtlik \etal 1998, Khrykin \etal 2017 )
in the \lya forest surrounding quasars. We have also not considered
the effect of stochasticity, as in our modeling the \lya emission
is deterministically related to the matter density.

\subsection*{Acknowledgments}
RACC thanks Simon White for conversations which lead to 
work in this paper.
RACC is supported by NASA ATP grant NNX17AK56G, NSF AST-1615940, and NSF AST-1614853.

Funding for SDSS-III has been provided by the Alfred P. Sloan Foundation, the Participating Institutions, the National Science Foundation, and the U.S. Department of Energy Office of Science. The SDSS-III web site is http://www.sdss3.org/.

SDSS-III is managed by the Astrophysical Research Consortium for the Participating Institutions of the SDSS-III Collaboration including the University of Arizona, the Brazilian Participation Group, Brookhaven National Laboratory, Carnegie Mellon University, University of Florida, the French Participation Group, the German Participation Group, Harvard University, the Instituto de Astrofisica de Canarias, the Michigan State/Notre Dame/JINA Participation Group, Johns Hopkins University, Lawrence Berkeley National Laboratory, Max Planck Institute for Astrophysics, Max Planck Institute for Extraterrestrial Physics, New Mexico State University, New York University, Ohio State University, Pennsylvania State University, University of Portsmouth, Princeton University, the Spanish Participation Group, University of Tokyo, University of Utah, Vanderbilt University, University of Virginia, University of Washington, and Yale University.

\section*{Appendix A: Light contamination}

The light from all one thousand fibres in the spectrograph is
dispersed onto the same 4096 column CCD. There is therefore
the potential for light from one fibre to leak into the 
extraction aperture for another fibre. The 
data reduction pipeline (Bolton et al. 2012)  has been designed so that this
level of light contamination is negligible for almost all purposes.
In carrying out IM with the LRG spectra, we are however operating
beyond what the instrument was designed to do. In C16 it was shown that 
for a given galaxy, light from quasars within 4 fibres measurably
contaminates the galaxy spectrum. 
The effect of the contamination is such that each pixel receives a small
fraction of the light from the same wavelength pixel of the contaminating
spectrum.
In Table \ref{fcontam} we show
how this fraction, $f_{\rm contam}$ depends on fibre separation,
 $\Delta_{\rm fibre}$.

\begin{table}
\centering
\caption{The fraction of light contaminating nearby fibres as a 
function of fibre separation on the CCD}
\label{fcontam}
\begin{tabular}{ll}
$\Delta_{\rm fibre}$ & $f_{\rm contam}$                  \\
1                     & $2\times10^{-3}$   \\
2                     & $6.5\times10^{-4}$ \\
3                     & $2\times10^{-4}$   \\
4                     & $2\times10^{-4}$  
\end{tabular}
\end{table}

\begin{figure}
\centerline{
\includegraphics[width=0.45\textwidth]{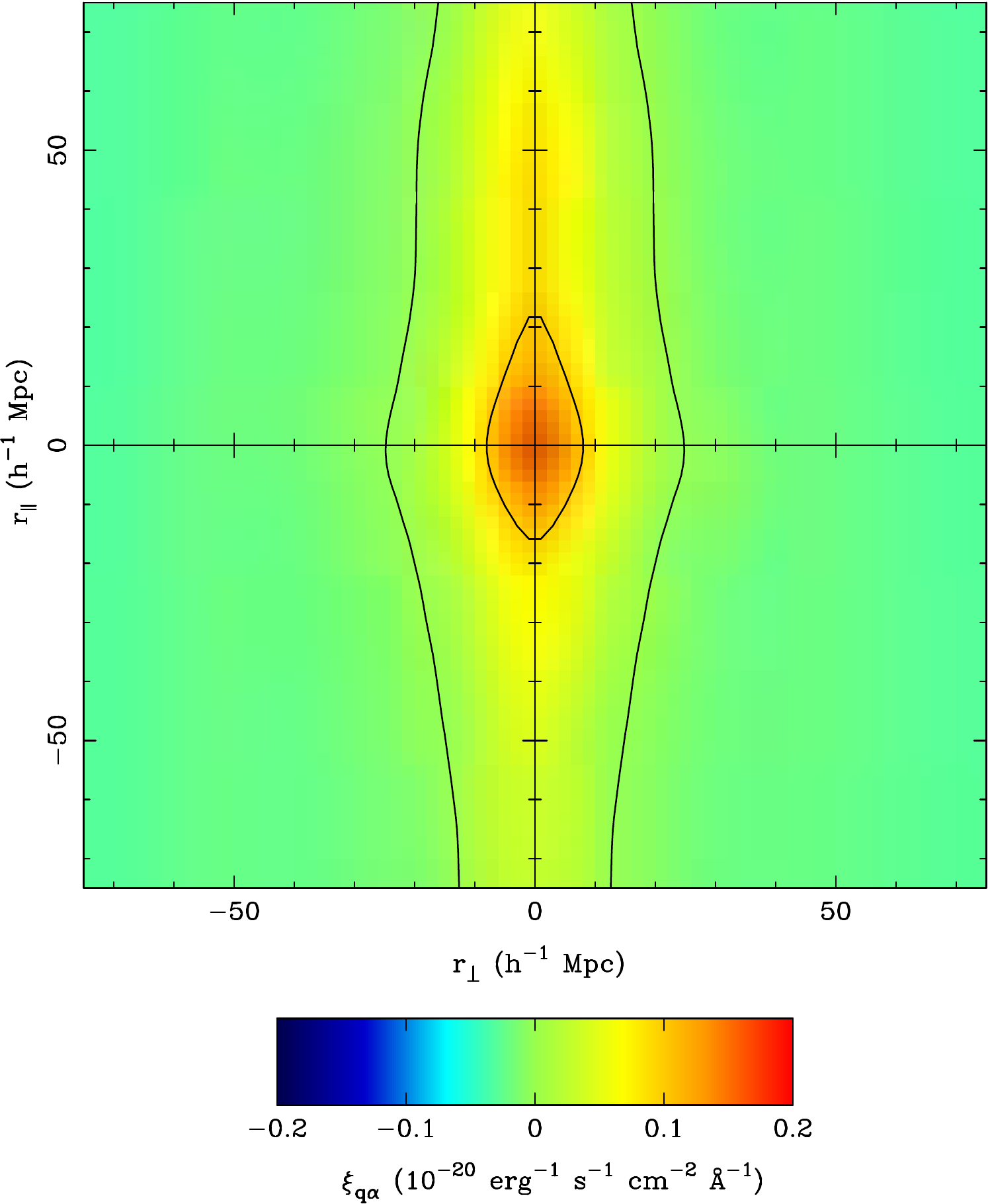}
}
\caption{ Light contamination: The cross-correlation function of quasars
and  \lya emission, $\xi_{q\alpha}$ from mock spectra generated
using light contamination only (see Section {mockfibre})).
\label{qecorrcontam}}
\end{figure}

In C16 the contamination was dealt with during the computation of
the cross-correlation function of quasars and \lya\ emission. Pairs
of quasars and \lya\ emission pixels
with $\Delta_{\rm fibre}$ equal to 5 or less
were excluded from the computation. 
Unfortunately we have discovered in the present paper that this was not
sufficient to remove all effects of contamination from \xiqe. This is
because additional contamination enters at second order, due to quasar
clustering, as follows. Suppose we exclude a particular quasar-pixel pair from 
the \xiqe computation. Because quasars are strongly clustered, there is a 
chance that another quasar is near to the excluded one and light from that 
quasar is also contaminating the pixel. The likelihood  of this contamination
occuring will depend on the clustering strength of quasars. This possibility
was ignored during the calculations in C16, but here we find
that it should be dealt with if the cross-correlation
results are to be reliable.

\begin{figure}
\centerline{
\includegraphics[width=0.45\textwidth]{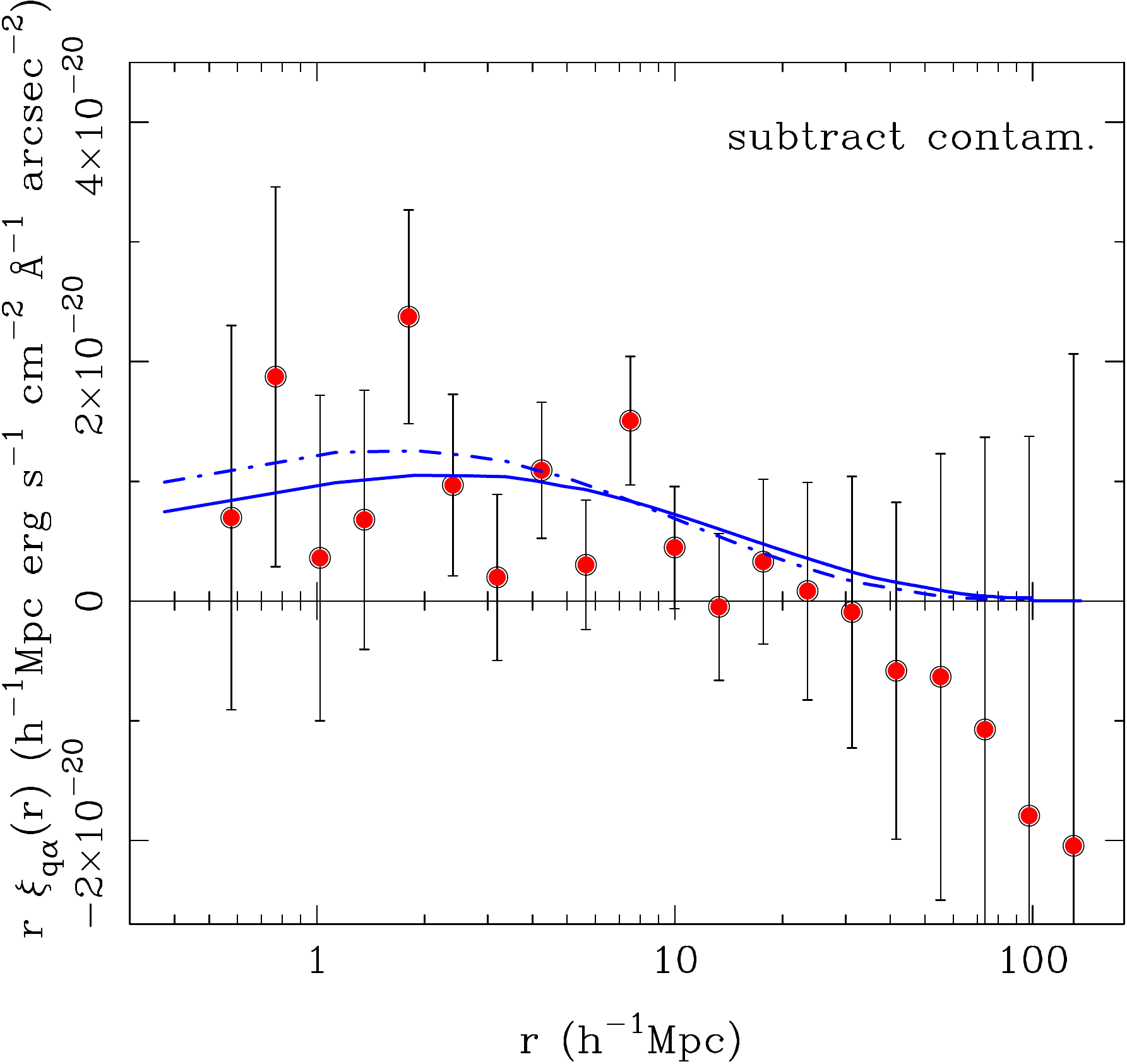}
}
\caption{ The cross-correlation function of quasars
and  \lya emission, $\xi_{q\alpha}$. We have removed 
light contamination by using decontamination method (3) (see text): 
 subtraction of $\xi_{q\alpha}$ 
computed from mock spectra generated
using light contamination only (see Section \ref{mockfibre}).
\label{qecorrcontam}}
\end{figure}

\subsection{Modelling contamination in \xiqe}
\label{mockfibre}

We model the effects of light contamination on \xiqe by making 
mock LRG fibre datasets which include only the light leakage from
quasars, but not the light from the LRGs themselves. Measuring \xiqe from 
these mock LRG spectra means that any \xiqe signal seen will be from
light contamination.

\begin{figure*}
\centerline{
\includegraphics[width=0.9\textwidth]{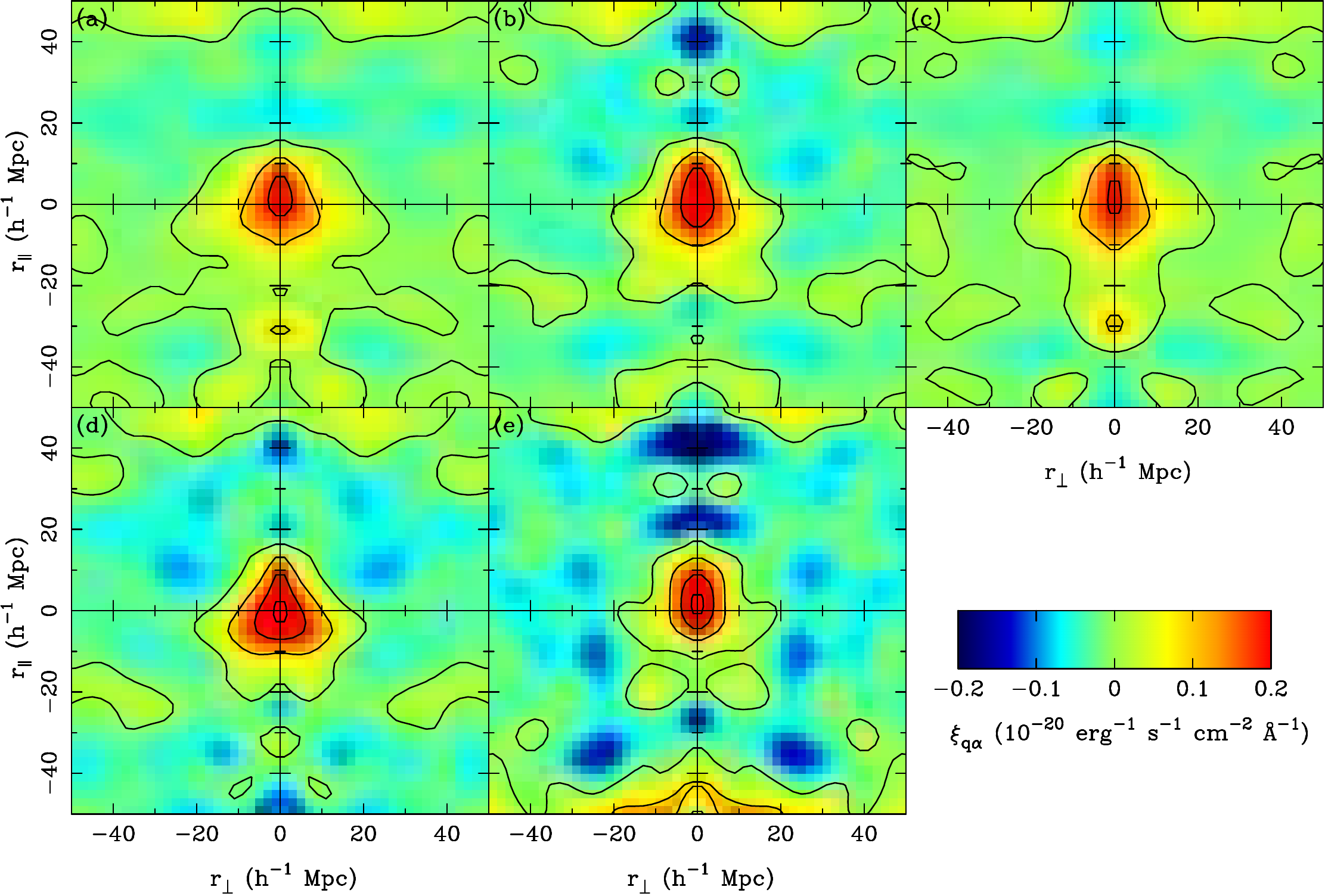}
}
\caption{ The quasar-\lya emission cross-correlation function,                 \xiqar\ (see Equation \ref{xieq}) as a function of quasar-pixel
separation across and along the line of sight.
The different panels represent different ways of dealing with 
contamination or splitting the data.
(a) The fiducial result (decontamination method 2), vetoing all pixels from 
the dataset which are within $\deltafibre=5$  and 75 mpc/h in the 
$r_{\parallel}$
direction from a quasar. (b) All pixels in spectra within
$\deltafibre=5$  of a quasar are excised from the dataset (decontamination method 1).
(c) The contamination modeled in Section \ref{mockfibre} is subtracted from 
the cross-correlation function (decontamination method 3). (d) Only pixels in spectra from the
center of the field of view are used in the (otherwise fiducial)
dataset. (e) Only pixels in spectra from the edge  of the field of view are 
used in the (otherwise fiducial)                
dataset.
  \label{sptest}}
\end{figure*}

To make the mock datasets we use the
information from Table \ref{fcontam}. For each LRG fibre we find any quasars
within  $\Delta_{\rm fibre}=4$ or less, and then add a fraction of 
the light in a quasar spectrum (given in Table \ref{fcontam}) to the 
LRG spectrum. We have tried two different approaches
for doing this. In the first,
we use the stacked mean quasar spectrum (taken from Section A2 in C16)
as our contaminating light, using the same spectrum for every galaxy fibre.
In the second case, we have used the actual SDSS quasar spectrum for the 
quasar in question as the contaminating light (again scaled appropriately).
We find no significant difference in the \xiqe results from the 
two methods. This is not the case however when looking at the
 \lya\ forest- \lya\ emission cross-correlation
(Section \ref{xifecontam}), where the stacked spectrum
does not exhibit the individual \lya\ forest fluctuations which cause 
contamination.

Once the mock LRG spectra have been made, we compute \xiqe from them
(Equation \ref{xieq}, using $w_{ri}=1$). When doing this, we 
exclude pairs of quasars and pixels separated by 5 fibres or less,
so that any contamination signal seen will be due to quasar
clustering .
The results are shown in Figure \ref{qecorrcontam},
where we plot \xiqe as a function of distance across and along the line
of sight. We can see that the light contamination does result in
a signifcant \xiqe signal, with the influence stretching along
the line of sight. Comparing Figure \ref{qecorrcontam} to Figure
\ref{sigpi} (which shows \xiqe without contamination), we can see that the
true signal is more centrally concentrated, dominating over the contamination,
on scales $|r_{\perp} < 10 \hmpc$ and $|r_{\parallel}| < 10 \hmpc$.
The contamination is however substantial,  reaching 50\% of the signal value
 even on these scales. It is therefore important to robustly 
remove it.

\subsection{Removing contamination from \xiqe}
\label{removecontam}
We carry out removal of this extra light contamination using three
different approaches. We test that they give consistent results for
\xiqe below.

(i) Decontamination method (1): Removal of all galaxy fibres within
$\Delta_{\rm fibre}=5$ from the dataset. This is the most extreme
solution- if any LRG fibre is within 5 fibres or less of a QSO fibre it
is completely removed from the dataset. Unfortunately, the significant
number of quasar fibres on each plate mean that the majority of
galaxy fibres is removed in this method. 
Of 1570095 LRG fibres in the initial
dataset, only 574603 (37 \%) remain after culling. This greatly reduces the
statistical power of  the measurement.

(ii) Decontamination method (2): Similar to method (1), we remove
contaminated LRG pixels from the dataset. However instead of removing an
entire fibre if it is within $\deltafibre =5$ or less of a QSO, we excise data
on a pixel by pixel basis. If the pixels are within the $\deltafibre$
constraint we remove them,  but only if they are also within $\pm75 \hmpc$
 in the line-of-sight direction of any  quasar. This deals with 
contamination, but only 8.4 \% of the pixels are removed. This 
decontamination method is our fiducial one, and was used to compute the
results shown in Figure \ref{cdmfid} and \ref{sigpi}.

(iii) Decontamination method (3): We compute \xiqe using the entire
dataset, except for quasar-pixel pairs within $\deltafibre=5$ (as was done
in C16). We then subtract \xiqe computed from contamination alone, using
the mock LRG spectra of Section \ref{mockfibre} 
(i.e. we subtract what is plotted in 
Figure \ref{qecorrcontam}).

\begin{figure}
\centerline{
\includegraphics[width=0.45\textwidth]{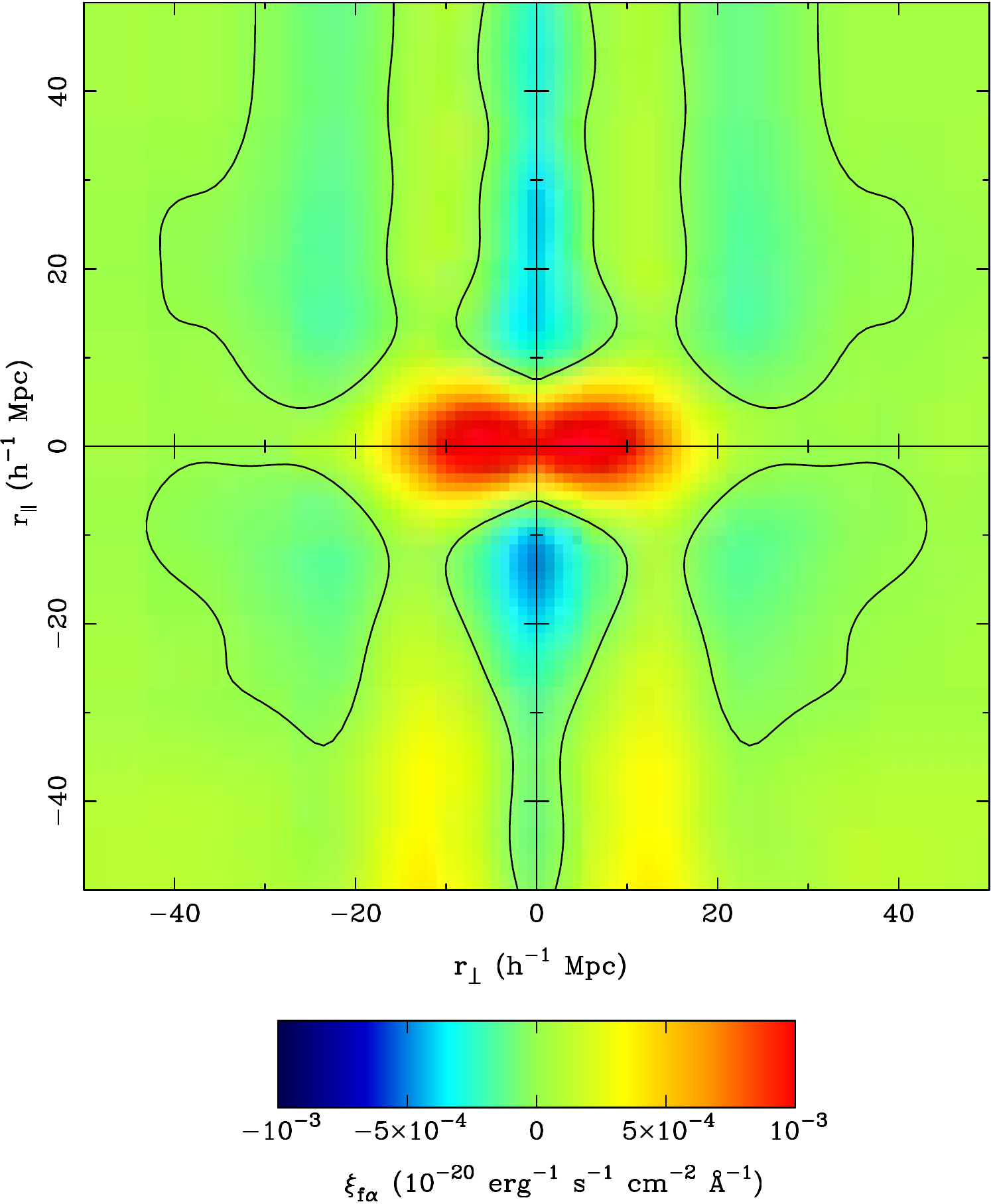}
}
\caption{ Light contamination: The cross-correlation function of \lya emission and the \lya
forest, $\xi_{f\alpha}$ for spectra from  BOSS DR12 (points with error bars).
  The actual signal predicted for $\xi_{f\alpha}$
in real data is negative (because the \lya forest has negative bias). 
\label{fecorrcontam}}
\end{figure}

In Figure \ref{sptest}, we show the \xiqe results for BOSS DR12 after applying
 our decontamination methods (the three panels in the top row). Each of the
panels is qualitatively similar which is not a trivial result,
 given the large differences
in how the contamination is dealt with by the three methods. All 
panels exhibit \lya emission centered on the origin,
with similar amplitude, and with a hint of elongation in the line of sight
direction. To determine the level of quantitative agreement, we fit the 
linear CDM model of Equation \ref{model}. The amplitude, 
\ampfa for method (1)
is $\text{\ampqa}  = 1.21^{+0.49}_{-0.49} \times 10^{-20} \ergs \cm^{-2} \angs^{-1} \asec^{-2}$,
for method (2), 
 $\text{\ampqa}  = 1.51^{+0.31}_{-0.29} \times 10^{-20} \ergs \cm^{-2} \angs^{-1} \asec^{-2}$,
and for
method(3) is
 $\text{\ampqa}  = 1.45^{+0.29}_{-0.32} \times 10^{-20} \ergs \cm^{-2} \angs^{-1} \asec^{-2}$,
These amplitude
results are all within one sigma of each 
other (as are the shape parameter $\Omega$ fits). These results give us 
confidence that the contamination due to quasar clustering identified 
above has been dealt with.

In C16, evidence was found that the fibres on the edge of
the telescope field of view (where the camera image has lower
optical quality) yield a smaller signal (by $2 \sigma$
compared to the fiducial results in that paper) than those in 
the center of the field of view.  We have repeated this test with the
BOSS DR12 dataset with the \xiqe results being shown in the bottom panel of
Figure \ref{sptest}. One can see that the pattern in 
$r_{\parallel}-r_{\perp}$
space is similar to the upper panels. Although panel (e) 
(edge fibres) appears to have a large amplitude, the statistical
error bar is
also significantly larger (the large fluctuations  seen 
away from the origin are also due to this). 
A fit to the CDM model for the center and edge fibres yields
an amplitude of 
 $\text{\ampqa}  = 1.22^{+0.36}_{-0.34} \times 10^{-20} \ergs \cm^{-2} \angs^{-1} \asec^{-2}$,
 and 
 $\text{\ampqa}  = 2.10^{+0.74}_{-0.70} \times 10^{-20} \ergs \cm^{-2} \angs^{-1} \asec^{-2}$,
 respectively.
This is consistent with 
what was seen in C16, that the reduction in optical quality at the edge
of the field of view make a signal detection more statistically uncertain.

We have seen that the contamination induced by quasar clustering can
be removed. It
is of course possible that some other type of unidentified contamination 
is responsible for the remainder of what we see in the \xiqe results. Two
pieces of evidence supporting the fact that it is actual \lya\ emission
are the consistency with the extrapolation of results from smaller scales
(Figure \ref{proj}), and also consistency with a simple physical
model which also explains the \xife results (Section \ref{secsim}).

\begin{figure}
\centerline{
\includegraphics[width=0.45\textwidth]{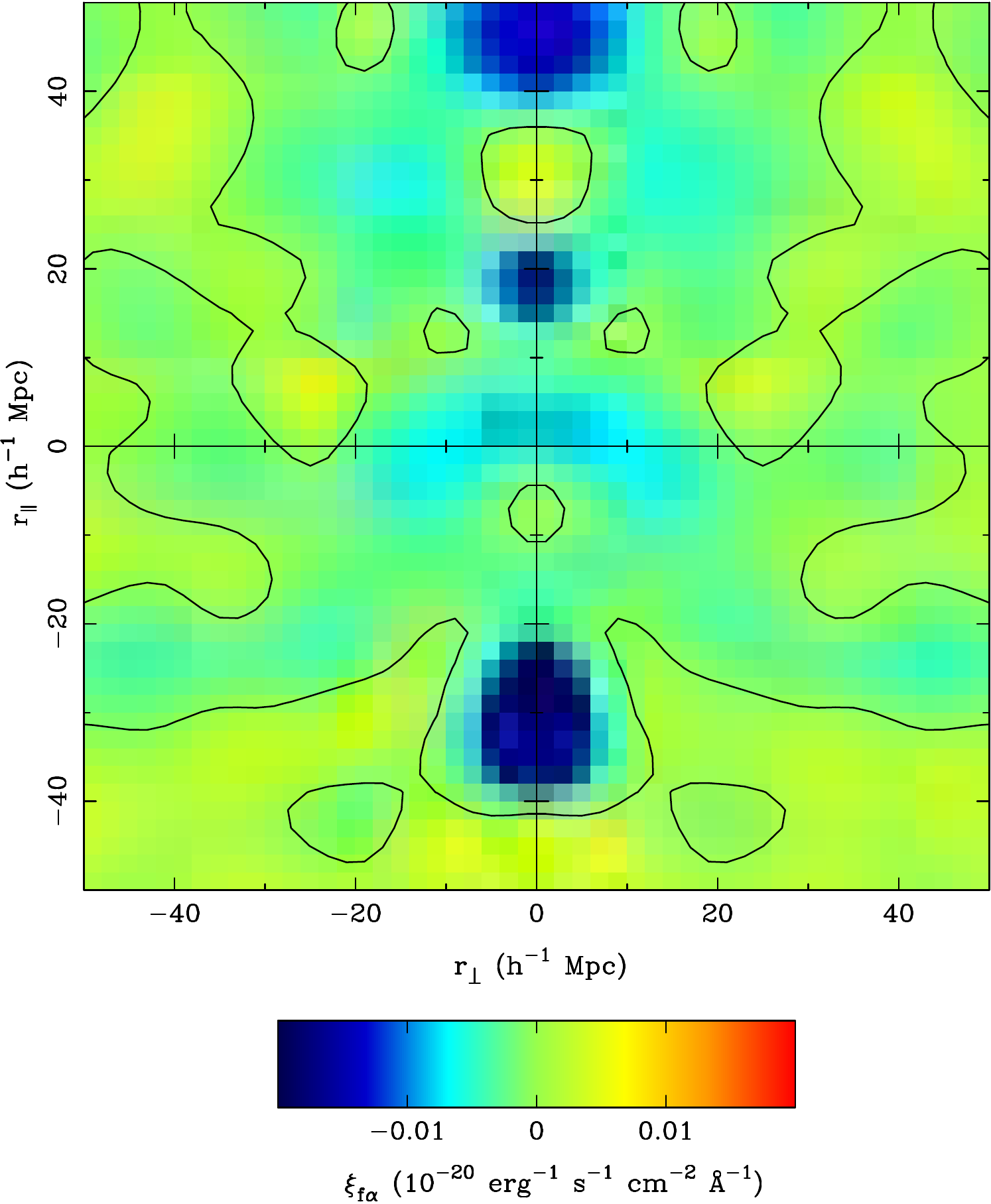}
}
\caption{ The cross-correlation function of \lya emission and the \lya
forest, $\xi_{f\alpha}$ for spectra from  BOSS DR12 (points with error bars)
as a function of separation parallel to and perpendicular to the line 
of sight.
We have eliminated all fibres within $\deltafibre=\pm5$ of a quasar before
computing the cross-correlation.
  The predicted amplitude of  $\xi_{f\alpha}$
is negative (because the \lya forest has negative bias). The cross-correlation
function plotted here is consistent with noise.
\label{fexispelim}}
\end{figure}

\subsection{Modelling contamination in \xife}
As quasar light contaminating galaxy fibres can yield a false 
\xiqe signal, we check whether this also affects the \lya\ forest-\lya\ 
emission cross-correlation. We expect that in this case, the contaminating 
quasar light will contain the same \lya\ forest fluctuations that are
being cross-correlated against, and so this will yield a false signal.
We model the contamination using our mock LRG fibre spectra datasets 
from Section \ref{mockfibre}. As mentioned above, we use the individual
quasar spectra as contaminants rather than the stacked spectrum, in order
that the \lya\ forest fluctuations be present in the contamination. 
Results for \xife from these mock spectra are shown in Figure 
\ref{fecorrcontam}. When comparing with the \xife results from the DR12
data (Figure \ref{fexisp}) it is important to realise that the color
scales are very different (by a factor of 20 on the scale bar)
, and that the contamination seen in Figure \ref{fecorrcontam} is quite
small. It is nevertheless significant, however, and we can see that 
the cross-correlation  of the negative bias \lya forest
fluctuations with the contaminating \lya forest fluctuations (also
with negative bias)
in the mock \lya emission spectra leads to a \xife signal which is 
positive. This is the opposite sign to that expected from the genuine
forest-emission correlation.

\label{xifecontam}

\subsection{Removing contamination from \xife}
The contamination can be dealt with in a similar fashion as for
\xiqe. Unlike the case of cross-correlating with quasars, however, there
is no easy way to apply method (2), because the objects to be 
cross-correlated with (the \lya\ forest pixels) fill the spectrum
(at least the interesting region where we are also looking 
for \lya\ emission). We therefore use two of 
 the three methods outlined in Section
\ref{removecontam}. Using Method (1) (excising whole fibres)
again results in a much smaller dataset, and higher statistical noise
in \xife. In Figure \ref{fexispelim} we show the \xife results using 
method (1). These can be compared to Figure \ref{fexi} in the main text,
which uses subtraction of contamination (method (3)). 
Figure \ref{fexispelim} is noticeably noisier. Both methods yield a 
non-detection of any \lya\ forest-emission cross-correlation.

We also try the edge/center test which was carried out with \xiqe. We
find that for fibres at the edge of the field-of-view the amplitude
of \xife clustering is 
$\text{\ampfa}  = (-5.2\pm{3.7}) \times 10^{-22} \ergs \cm^{-2} \angs^{-1}
\asec^{-2},$                       
 For fibres at the center
we find 
$\text{\ampfa}  = -2.5_{-2.9}^{+1.8} \times 10^{-22} \ergs \cm^{-2} \angs^{-1}
\asec^{-2}$. Both are consistent with no
measurable forest-emission correlation.


\begin{thebibliography}{}
\bibitem[Ahn et al.(2014)]{Ahn14} 
Ahn, C., \etal 2014, \apjs, 211, 17

\bibitem[]{}
Alam, S., 2015, ApJS, 219, 12

\bibitem[]{}
Albareti et al., 2017, ApJS, 233, 25


Bajtlik, S., Duncan, R. C., Ostriker, J. P., 1988, ApJ 327, 570

\bibitem[]{}
Bandura, K. et al., 2014, Proc.
SPIE Int. Soc. Opt. Eng., 9145:22, 2014. doi: 10.1117/12.2054950.



\bibitem[]{}
Benitez, N., et al., 2014, J-PAS Red Book, arXiv:1403.5237

\bibitem[Blanc et al.(2011)]{Blanc11} 
Blanc, G.A., \etal, 2011, 736, 31


\bibitem[Bolton et al.(2012)]{Bolton12}
Bolton, A., \etal 2012, \aj, 144, 144

\bibitem[]{}
Borisova, E., et al., 2016, ApJ, 831, 39

\bibitem[Bouwens et al.(2010)]{Bouwens10}
Bouwens, R.~J., \etal 2010, \apj, 709, L133

\bibitem[Bovy et al.(2011)]{Bovy11}
Bovy, J., \etal 2011, \apj, 729, 141


\bibitem[]{}
Busca, N. et al., 2013, A \& A, 552, 96


\bibitem[]{}
Cantalupo, S., Arrigoni-Battaia, F., Prochaska, J. X., Hennawi, J. F., \&
Madau, P. 2014, Nature, 506, 63

\bibitem[]{}
Carilli,C.L., 2011, ApJ Lett., 730, L30


\bibitem[Cassata et al.(2011)]{Cassata11}
Cassata, P. \etal 2011, \aa, 525, 143

\bibitem[]{}
Castander, F.J., \etal 2012, Proc. SPIE, 8446, 6

\bibitem[]{}
Chang, T.-C., Pen, U.-L., Bandura, K., and Peterson, J. B., 2010, Nature,
466, 463

\bibitem[]{}
Cisewski, J,. Croft, R.A.C., Freeman, P.,E., Genovese, C.R., Khandai, N.,
Ozbek, M., and Wasserman, L., 2014, MNRAS, 440, 2599

\bibitem[Dawson et al.(2013)]{Dawson13}
Dawson, K.S., \etal, 2013, \aj, 145, 10


\bibitem[]{}
Davis, M., Newman, J. A., Faber, S. M. \& Phillips, A. C. 2001, in Deep Fields (eds Cristiani, S. Renzini, A. \& Williams, R. E.) 241 (Springer)

\bibitem[]{}
Delubac, T., \etal 2015, 574, 59

\bibitem[]{}
Di Matteo, T., Khandai, N., DeGraf, C., Feng, Y., Croft, R. A. C.,
 Lopez, J., and Springel, V., 2012, ApJ, 745, 29

\bibitem[]{}
Dor\'{e} \etal 2014, arXiv:1412.4872

\bibitem[]{}
Eisenstein, D.J, et al, 2011, Astronomical Journal 142, 72

\bibitem[Font-Ribera et al.(2013)]{Font-Ribera13}
Font-Ribera, A., \etal 2013, JCAP, 05, 018


\bibitem[Font-Ribera et al. (2014)]{Font 2014}
Font-Ribera, A, et al., 2014, JCAP, 05,27

\bibitem[Gallego]{}
Gallego, S.G. et al., 2018, MNRAS, 475, 3854

\bibitem[Gawiser et al.(2007)]{Gawiser07}
Gawiser, E., \etal 2007, \apj, 671, 278


\bibitem[]{}
Gould, A., \& Weinberg, D. H. 1996, ApJ, 468, 462

\bibitem[Gronwall et al.(2007)]{Gronwall07}
Gronwall, C., \etal, 2007, \apj, 667, 79

\bibitem[Guaita et al.(2010)]{Guaita10}
Guaita, L., \etal 2010, \apj, 714, 255

\bibitem[Gunn et al. (1998)]{Gunn98}
Gunn, J.E., et al. 1998, AJ, 116, 3040

\bibitem[Gunn et al. (2006)]{Gunn06}
Gunn, J.E., et al. 2006, AJ, 131, 2332


\bibitem[]{}
Hennawi, J. F., Prochaska, J. X., Cantalupo, S., \& Arrigoni-Battaia, F. 2015,
Sci, 348, 779


\bibitem[]{}
Hernquist, L., Katz, N., Weinberg, D. H. and Miralda-Escud\'{e}, J.,
ApJ,  457, L51

\bibitem[]{} 
Hill, G.J. \etal, 2016,
Proceeding of "Multi-Object Spectroscopy in the Next Decade", Eds.  Skillen, I., Balcells, M., \& Trager S., ASP Conference Series, Vol. 507. San Francisco: Astronomical Society of the Pacific, 2016, p.39


\bibitem[]{}
Hogan, C. J., \& Weymann, R. J. 1987, MNRAS, 225, 1

\bibitem[Kaiser(1987)]{Kaiser87}
Kaiser, N. 1987, \mnras, 227, 1

\bibitem[]{}
Kakiichi, K and Dijkstra, M., 2017, MNRAS, submitted, arXiv:1710.10053

\bibitem[]{}
Keating, G.K., et al., (2015), ApJ, 814, 140	

\bibitem[Kirkpatrick et al.(2011)]{Kirkpatrick11}
Kirkpatrick, J.A., Schlegel, D.J., Ross, N.P.,
Meyers, A.D., Hennawi,J.F., Sheldon, E.S., Schneider, D.P.,
Weaver, B.A., 2011, \apj, 743, 125


\bibitem[]{}
Kollmeier, J. A., Zheng, Z., Davé, R., et al. 2010, ApJ, 708, 1048

\bibitem[]{}
Kovetz, E.D., et al., 2017, Physics Reports, submitted, arXiv:1709.09066

\bibitem[]{}
Khrykin, I. S., Hennawi, J. F.\& McQuinn, M., 2017, ApJ, 838, 96

\bibitem[]{}
Lee, K.G., \etal 2013, AJ, 145, 69

\bibitem[]{}
Madau, P., Meiksin, A., \& Rees, M.J., ApJ 475, 429 1997, 

\bibitem[]{}
Martin, D. C., Chang, D., Matuszewski, M., et al. 2014, ApJ, 786, 106

\bibitem[Matsuda et al.(2012)]{Matsuda12}
Matsuda, Y., Yamada,
T., Hayashino, T., et al.\ 2012, \mnras, 425, 878


\bibitem[]{}
McDonald, P., 2003, ApJ, 585, 34

\bibitem[Momose et al.(2014)]{Momose14} Momose, R., Ouchi, M.,
Nakajima, K., et al.\ 2014, \mnras, 442, 110

\bibitem[]{}
Noterdaeme, P, Petijean, P., Ledoux, C., \& Srianand, R. , 2009, A\&A, 505, 1087

\bibitem[]{}
Noterdaeme, P, et al., 2012,  A\&A, 547,1 


\bibitem[Ouchi et al.(2008)]{Ouchi08}
Ouchi, M., \etal 2008, \apjs, 176, 301


\bibitem[]{}
Ozbek,M., and Croft, R.A.C., 2016. MNRAS, 456, 3610

\bibitem[P\^{a}ris et al.(2012)]{Paris12}
P\^{a}ris, I. \etal 2012, \aa, 548, 66


\bibitem[]{}
P\^{a}ris, I., \etal, 2017, A\&A, 597, 79 

\bibitem[Partridge \& Peebles(1967)]{Partridge67}
Partridge, R.B. \& Peebles, P.J.E., 1967, \apj, 147, 868

\bibitem[Peeples et al. (2010)]{peeples10} 
Peeples, M.S., Weinberg, D.H,
Dav\'{e}, R., Fardal, M., \& Katz N., 2010, \mnras 401, 1281

\bibitem[]{}
Pullen, A.R., Dor\'{e}, O., and Bock, J.,  2014, ApJ, 786, 111

\bibitem[]{}
Pullen, A.R., Hirata, C., Dor\'{e}, O. and Racanelli, A.  2016, PASJ, 68, 12


\bibitem[]{}
Pullen, A.R., Serra, P., Chang, T.-C., Dor\'{e}, O., and Ho, S. 2018, MNRAS,{\it tmp} 1181


\bibitem[]{}
Ross, N.P., et al., 2013, ApJ, 773, 14

\bibitem[]{}
Schlegel,D.J., Finkbeiner, D.P. \& Davis, M., 1998, ApJ, 500, 525 

\bibitem[Silva et al.(2013)]{Silva13}
Silva, M., Santos, M., Gong, Y. And Cooray, A., 2013, \apj, 763, 132


\bibitem[]{}
Slosar, A., et al. 2013, JCAP, 04, 026

\bibitem[Smee et al.(2013)]{Smee13}
Smee, S.A., \etal 2013, \aj ,126, 32


\bibitem[]{}
Smith, A., Tsang, B.T.H., Bromm, V., \& Milosavljević, Miloš

\bibitem[]{}
Springel, V. and Hernquist L., 2003, MNRAS, 339, 289

\bibitem[]{}
Springel, V. 2005, MNRAS, 364, 1105

\bibitem[Steidel et al.(2011)]{Steidel11}
{Steidel}, C.~C., {Bogosavljevi{\'c}}, M., {Shapley}, A.~E.,
{Kollmeier}, J.~A., {Reddy}, N.~A., {Erb}, D.~K. and {Pettini}, M.
\apj, 2011, 736, 160

\bibitem[]{}
Wold, I. G. B., Finkelstein, S. L., Barger, A. J., Cowie, L. L.\& Rosenwasser, B.,
ApJ 848, 108

\bibitem[]{}
Y\'eche, C., et al., 2010, A\&A, 523, 14

\bibitem[]{}
Zheng,Z., Cen, R., Trac, H., Miralda-Escud\'{e}, J.,
2010, \apj, 716, 574


\bibitem[Zheng et al.(2011a)]{Zheng11a}
Zheng,Z., Cen, R., Trac, H., Miralda-Escud\'{e}, J.,
2011, \apj, 726, 38


\end{thebibliography}
\end{document}